\documentclass[11pt,reqno]{amsart}

\usepackage{tikz}
\usetikzlibrary{arrows.meta,decorations.markings}
\usepackage[T1]{fontenc}
\usepackage{lmodern}
\usepackage{microtype}
\usepackage{mathtools}
\usepackage{amssymb,enumitem}
\usepackage{bbm}
\usepackage{amsthm}
\usepackage{xcolor}

\usepackage{orcidlink}

\usepackage{hyperref}

\hypersetup{
	colorlinks=true,
	linkcolor=blue!55!black,
	citecolor=blue!55!black,
	urlcolor=blue!55!black
}

\definecolor{green}{rgb}{0.0, 0.5, 0.5}

\definecolor{lgray}{gray}{0.9}
\definecolor{llgray}{gray}{0.95}
\definecolor{lllgray}{gray}{0.975}

\allowdisplaybreaks
\numberwithin{equation}{section}

\newtheorem{theorem}{Theorem}[section]

\newtheorem{lemma}[theorem]{Lemma}
\newtheorem{corollary}[theorem]{Corollary}
\theoremstyle{definition}
\newtheorem{assumption}[theorem]{Assumption}
\newtheorem{definition}[theorem]{Definition}
\theoremstyle{remark}
\newtheorem{remark}[theorem]{Remark}

\newcommand{\A}{\mathcal A}
\newcommand{\F}{\mathcal F}

\newcommand{\one}{{\let\mathbb\mathbbm\mathbb{1}_{\mathcal F}}}
\newcommand{\dd}{\mathrm d}
\newcommand{\iu}{\mathrm i}
\newcommand{\e}{\mathrm e}
\newcommand{\Tr}{\operatorname{Tr}}
\newcommand{\dist}{\operatorname{dist}}

\newcommand{\norm}[1]{\left\lVert #1\right\rVert}
\newcommand{\abs}[1]{\left\lvert #1\right\rvert}
\newcommand{\ip}[2]{\left\langle #1,#2\right\rangle}
\newcommand{\ac}[2]{\left\{#1,#2\right\}}

\newcommand{\tauFH}{\tau^{\mathrm{FH}}}
\newcommand{\tauf}{\tau^{\mathrm f}}

\tikzset{
	lattice picture/.style={
		x=1.4cm,
		y=1.4cm
	},
	lattice grid/.style={
		draw=gray!35,
		line width=0.4pt
	},
	shortest path/.style={
		line width=1.3pt,
		line cap=round,
		line join=round
	},
	red path/.style={
		shortest path,
		draw=red
	},
	blue path/.style={
		shortest path,
		draw=blue
	},
	blue dark/.style={
		shortest path,
		draw=blue!80!black
	},
	blue medium/.style={
		shortest path,
		draw=blue!70
	},
	blue light/.style={
		shortest path,
		draw=blue!45
	},
	path arrows/.style n args={2}{
		postaction={decorate},
		decoration={
			markings,
			mark=between positions #1 and 1 step #2 with {
				\arrow{Stealth[length=1.8mm,width=1.3mm]}
			}
		}
	}
}

\newcommand{\latticebackground}[1]{%
	\fill[gray!7] (0,0) rectangle (#1,1);
	\draw[lattice grid, step=1] (0,0) grid (#1,1);
}

\newcommand{\latticevertices}[1]{%
	\foreach \x in {0,...,#1}{
		\foreach \y in {0,1}{
			\fill[black!70] (\x,\y) circle[radius=1.7pt];
		}
	}
}

\title[Comparing weakly interacting and free dynamics]{Comparing weakly interacting and free quantum dynamics locally}

\author[J.~A.~Flores]{Jorge Acu{\~n}a Flores\,\orcidlink{0009-0004-9023-0615}}
\address{Department of Mathematics, University of T\"ubingen, Auf der Morgenstelle 10,
	72076 T\"ubingen, Germany}
\email{jorge.acuna-flores@uni-tuebingen.de}

\author[J.~Henheik]{Joscha Henheik\,\orcidlink{0000-0003-1106-327X}}
\address{Mathematical Institute, University of Oxford, Andrew Wiles Building, Radcliffe Observatory Quarter, Woodstock Road, Oxford OX2 6GG, United Kingdom}
\email{joscha.henheik@maths.ox.ac.uk}

\author[M.~Lemm]{Marius Lemm\,\orcidlink{0000-0001-6459-8046}}
\address{Department of Mathematics, University of T\"ubingen, Auf der Morgenstelle 10,
	72076 T\"ubingen, Germany}
\email{marius.lemm@uni-tuebingen.de}

\author[D.~Toniolo]{Daniele Toniolo\,\orcidlink{0000-0003-2517-0770}}
\address{Department of Mathematics, University of T\"ubingen, Auf der Morgenstelle 10,
	72076 T\"ubingen, Germany}
\email{daniele.toniolo@uni-tuebingen.de}

\begin{document}
	
	\begin{abstract}
		Lieb--Robinson bounds (LRBs) have become a standard tool for perturbative results in quantum many-body physics, including gap stability and many-body adiabatic theory. Here, we use LRBs to locally approximate weakly interacting many-body dynamics with one-body dynamics on large, perturbative time scales. This allows us  to use one-body transport results to establish the expected many-body transport behavior on finite time scales in the following two illustrative settings:
		(i) We prove quantum diffusion, as quantified by a certain operator spreading diagnostic, in weakly interacting and weakly disordered Fermi systems in dimensions $D\geq2$ on large finite time scales based on the non-interacting diffusion result \cite{BlackDroginHernandez2025}. 
		(ii) We show for Fermi--Hubbard models with initially low spin-down density, that all local multi-time spin-up correlation functions are approximated by their free counterparts. In particular, we obtain an OTOC (out-of-time-ordered correlator) counterpart to numerically observed patterns of short-time entanglement related to lattice path counts \cite{bippus2026two}.
		
	\end{abstract}
	\maketitle
	
	\section{Introduction and main results}
	Lieb--Robinson bounds \cite{Lieb_Robinson} have emerged as the decisive tool to perform volume-uniform perturbation theory in quantum lattice systems over the past twenty years. Major examples include the stability of spectral gaps \cite{michalakis2013stability,bravyi2010topological,Bravyi_Hastings_2010,nachtergaele2019quasi,nachtergaele2022quasi,del2025boundary}, many-body adiabatic theorems \cite{bachmann2018adiabatic,teufel2020non,henheik2022adiabatic}, and other prethermal phenomena \cite{Abanin_2017,mcdonough2025non}.

	Our goal in this paper is to exploit dynamical comparisons that follow from Lieb--Robinson bounds (LRBs) in settings \textit{without} typical additional assumptions such as a spectral gap or another parameter leading to slow dynamics. The LRB allows to approximate the Heisenberg evolution of a local observable living on a region $X$ at time $t$ by the Heisenberg evolution generated by a local Hamiltonian that lives on the region $X$ expanded by $r>vt$ with $v$ the Lieb-Robinson velocity. This  locality property immediately allows to perform perturbation theory inside the local region $X$ expanded by $vt$ that depends only on the time $t$ and not on the system size. This yields a remarkably general dynamical comparison principle for many-body dynamics on suitable perturbative time scales. 
	
	We present two concrete examples of this kind, in which we rigorously transfer information about quantum transport from tractable reference one-body dynamics to many-body dynamics with weak interactions, thereby confirming long-expected dynamical behavior on finite perturbative time scales.

	We recall the basic mechanism in the simplest setting where one uses the standard operator-norm LRB so that the idea is most easily visible. Consider a quantum many-body Hamiltonian on a finite graph \(\Lambda\subset \mathbb Z^D\) of the form
	\begin{equation}\label{eq:intro-H}
		H_\varepsilon=H_0+\varepsilon W,
		\qquad W=\sum_{\substack{x,y\in \Lambda:\\ x\sim y}} W_{xy}.
	\end{equation}
	Here, $H_0$ is a simpler reference Hamiltonian regarded as known (in fact, in this paper, $H_0$ will always be non-interacting), $x\sim y$ denotes nearest-neighbors and $\varepsilon\ll 1$ is a small parameter. Given a bounded local observable \(A\) supported at the origin, we write \[ \tau_t^\varepsilon(A)=e^{\iu tH_\varepsilon}Ae^{-\iu tH_\varepsilon}\] for the associated Heisenberg dynamics. The goal is to compare the perturbed many-body evolution $\tau_t^\varepsilon(A)$ with the one generated by $H_0$, $\tau_t^0(A)$.  The naive (global) comparison that follows from  Duhamel reads
	\begin{equation}\label{eq:intro-global-Duhamel}
		\norm{\tau_t^\varepsilon(A)-\tau_t^0(A)}
		\leq 2t\varepsilon  \norm W\norm A,\qquad t\geq 0.
	\end{equation}
	Since \(\varepsilon \norm W\) is proportional to the volume, \eqref{eq:intro-global-Duhamel} becomes useless in the thermodynamic limit because the norm of an extensive perturbation is, of course, a poor expansion parameter in an extended many-body system.  
	
	Instead, we consider the following three-step comparison procedure. 
	
	\begin{itemize}
		\item 
		\textit{(I) Initial localization:} Let $B_R=B_R(0)$ be a ball of radius $R>0$. By a standard application of Duhamel's formula, the Lieb-Robinson bound implies the local approximation
		\[
		\|\tau_t^{\varepsilon}(A)-\tau_t^{\varepsilon,B_R}(A)\|\lesssim \|A\| e^{vt-R},
		\]
		where $\tau_t^{\varepsilon,B_R}(A)=e^{\iu t H_\varepsilon\vert_{B_R}}A e^{-\iu t H_\varepsilon\vert_{B_R}}$ is the Heisenberg dynamics generated by $H_\varepsilon\vert_{B_R}$, the restriction of $H_\varepsilon$ to terms with supports inside $B_R$.
		\item 
		\textit{(II) Local comparison:} By a Duhamel argument, we remove the perturbation locally,
		\[
		\|\tau_t^{\varepsilon,B_R}(A)-\tau_t^{0,B_R}(A)\|\lesssim \|A\| \varepsilon t R^D
		\]
		where $\varepsilon R^D$ arises from the operator norm of the restricted perturbation $\varepsilon W\vert_{B_R}$ and the $t$-factor arises from the Duhamel time integration.
		\item \textit{(III) Extension (optional):} If the reference dynamics $\tau_t^{0,B_R}$ is generated by a $1$-body Hamiltonian, it is often convenient to extend it to the full space $\mathbb Z^D$ using one-body propagation bounds. This is essentially an inverse to Step (I).
		\[
		\|\tau_t^{0,B_R}(A)-\tau_t^{0,\mathbb Z^D}(A)\|\lesssim \|A\| e^{v't-R}
		\]
	\end{itemize}

	Combining steps (I)--(III) yields the comparison result
	\[
	\|\tau_t^{\varepsilon}(A)-\tau_t^{0,\mathbb Z^D}(A)\|\lesssim \|A\| \left(e^{vt-R}+e^{v't-R}+\varepsilon t R^D\right).
	\]
	Assuming $t\geq 1$, a  natural choice for the free parameter $R$ is then $R=ct$  with $c>\max\{v,v'\}$, so that the two exponentials decay rapidly and the error behaves as $\varepsilon t^{D+1}$. In other words, on time scales $t\leq \varepsilon^{-\alpha}$ for a suitable $\alpha>0$, the many-body dynamics is well-approximated by the simpler $H_0$-dynamics. (If the $H_0$-dynamics is still many-body, then one skips Step (III) extension to stay within the finite-but-large system formalism.)
	
	The fact that LRBs allow for such a simple perturbative procedure will not be surprising to experts. In fact, higher-order perturbation theory  has been used in similar ways to  study prethermal phenomena (see, e.g.,  \cite{Abanin_2017}), but we believe that its broad applicability has not been fully appreciated.
	In this paper, we propose to exploit such dynamical comparisons, implementing the general procedure (I)--(III), in two contexts where no gap (or related parameter leading to slow dynamics) is present:  the interacting Anderson model and Fermi--Hubbard Hamiltonians.
	
	\begin{itemize}
		\item 
		In Result 1, we show that an  operator-spreading version of the second moment of the position operator can be approximated by its non-interacting version. The operator-spreading diagnostic features a spatially growing weight $|x|^2$, which necessitates different locality bounds and eventually leads to the error bound $\varepsilon t^{D+3}$ (compared to the $\varepsilon t^{D+1}$ which is natural for local observables as described above). We then consider specifically the Anderson model as the non-interacting Hamiltonian, establishing quantum diffusion for the corresponding operator-spreading. 
		
		\item In Result 2, the small parameter is no longer in the Hamiltonian, but instead in the \emph{initial state}, which we assume to have a low density of spin-down particles. This clearly makes the perturbation step (II)  more subtle because operator norm LRBs are insensitive to the small parameter in the initial state. Instead, we exploit the perspectives of state-dependent LRBs that has recently been developed in the  context of lattice bosons for different reasons \cite{yin2022finite,faupin2022maximal,kuwahara2024effective,lemm2025quantum,faupin2025macroscopic,LRZ2026}.
	\end{itemize}
	
	These examples highlight that LRBs provide a remarkably simple  way of perturbatively understanding many-body dynamics  uniformly in the system size. This insight  has certainly been implicit in the locality bounds literature. Here, we wish to put this general principle at center stage to make it more widely known and to show its power in different settings than have previously been considered.
	
	We now summarize Results 1 \& 2 in more detail.
	
	\subsection{First Result: Weakly interacting fermions and quantum diffusion.}
	We consider $N$ spinless Fermions on a finite graph $\Lambda\subset \mathbb Z^D$. We study the spinless Fermion Hamiltonian
	\begin{equation}\label{eq:Hresult1intro}
		H_{\varepsilon}= H_0+\varepsilon
		\sum_{x,y\in\Lambda}
		U_{xy}n_xn_y,
		\qquad H_0=\sum_{x,y\in \Lambda} h_{xy}a_x^* a_y
	\end{equation}
	with the one-body part 
	\begin{equation}
		h=-\Delta+V
	\end{equation} a self-adjoint discrete Schr\"odinger operator on $\ell^2(\mathbb Z^D)$ and $h_\Lambda$ its open restriction in the definition of $H_0$. We emphasize that the interaction strength $\varepsilon\ll 1 $ is a small $\mathcal O(1)$ parameter, i.e., the system is still strongly interacting compared to, say, a mean-field scaling where $\varepsilon=\frac{1}{N}$ or other scaling limits coupling $\varepsilon$ to the system size. 
	
	\textit{General comparison result for second moments.} A key quantity in the analysis of the one-body dynamics is the second moment of the position operator 
	\begin{equation}
		\sum_{x\in\mathbb Z^D}|x|^2
		\abs{\ip{\delta_x}{\e^{\iu th}\delta_0}}^2
	\end{equation}
	which is a central object in studying one-body transport; see, e.g., \cite{damanik2007upper}. Here and below, $\one$ denotes the identity on the fermionic Fock space $\F_\Lambda:=\bigoplus_{n\geq0}\bigwedge^n\ell^2(\Lambda)$, and $|A|:=(A^*A)^{1/2}$ for $A\in\mathcal B(\F_\Lambda)$. Our first main result, Theorem~\ref{thm:transport-stability}, establishes that 
	\begin{equation}\label{eq:introexp}
		\begin{aligned}
			&\sum_{x\in \Lambda} |x|^2 \abs{\ac{\tau_{t}^{\varepsilon}(a_0^*)}{a_x}}^2\\
			&\quad= \left(\sum_{x\in\mathbb Z^D}|x|^2
			\abs{\ip{\delta_x}{\e^{\iu th}\delta_0}}^2\right)\one+
			\mathcal O\!\left(\varepsilon  t^{D+3}
			+t^{D+2}\e^{-ct}\right)
		\end{aligned}
	\end{equation}
	for $t\geq1$, with the error measured in operator norm. Here and below, the positive constant $c$ may change from line to line. We emphasize that this is an exceptionally strong approximation result: For \textit{any} state, the many-body operator $\abs{\ac{\tau_{t}^{\varepsilon}(a_0^*)}{a_x}}^2$ (which  measures the chance to observe, at time $t$ and position $x$,  the effect of adding a particle at the origin at time zero)  is close to a scalar multiple of the identity on the Fock space.

	The right-hand side of \eqref{eq:introexp} is the second moment of the probability measure $x\mapsto \abs{\ip{\delta_x}{\e^{\iu th}\delta_0}}^2$, in a time regime where the correction is small. One may therefore ask to what extent the operator on the left-hand side is normalized as well. A modification of the proof of \eqref{eq:introexp} yields the following normalized version of \eqref{eq:introexp}. Define
	\begin{equation} \label{eq:S_def}
		S_\varepsilon(t):=\sum_{y\in\Lambda}\abs{\ac{\tau_t^\varepsilon(a_0^*)}{a_y}}^2\in\mathcal B(\F_\Lambda).
	\end{equation}
	Analogously to \eqref{eq:introexp}, we also show in Theorem \ref{thm:transport-stability} the proximity of second moments, namely it holds:
	\begin{equation}\label{eq:introexp_normalized}
		\begin{aligned}
			&\qquad S_\varepsilon(t)^{-1/2}\left(\sum_{x\in \Lambda} |x|^2 \abs{\ac{\tau_{t}^{\varepsilon}(a_0^*)}{a_x}}^2\right)S_\varepsilon(t)^{-1/2}\\
			&\qquad= \left(\sum_{x\in\mathbb Z^D}|x|^2
			\abs{\ip{\delta_x}{\e^{\iu th}\delta_0}}^2\right)\one+
			\mathcal O\!\left(\varepsilon t^{D+3}
			+t^{D+2}\e^{-ct}\right)
		\end{aligned}
	\end{equation}
	where it is part of the proof that for sufficiently small values of $\varepsilon t^{D+3} +t^{D+2}\e^{-ct}$, $S_\varepsilon(t)^{-1}$  exists and is close to $\one$ up to errors $\mathcal O\!\left(\varepsilon t^{D+3} +t^{D+2}\e^{-ct}\right)$.

	The proofs of \eqref{eq:introexp} and \eqref{eq:introexp_normalized} follow exactly the three steps described above. 
	Equation \eqref{eq:introexp} holds for any external  potential $V$ in $h=-\Delta+V$. It allows to calculate the dynamical many-body  expression on the left-hand side of \eqref{eq:introexp} in terms of the second moment of the position operator of the one-body dynamics on the right-hand side for times smaller than a suitable power of $\varepsilon^{-1}$, where the power depends on the growth in $t$ of the first term. Notice that the reference one-body dynamics in \eqref{eq:introexp} is considered over all of $\mathbb Z^D$.\footnote{In fact, the approximation naturally compares the many-body dynamics on $\Lambda$ to the one-body dynamics on $\Lambda$, but we prefer to extend the one-body dynamics (which also satisfies a propagation bound) to $\mathbb Z^D$, because this is the setting where most results are proved. As a simple example, the free Laplacian is a simpler object on $\mathbb Z^D$ than on $\Lambda$, because it can be diagonalized by Fourier series.}

	\textit{Application to the interacting Anderson model.}
	This connection is of particular interest to us when the reference dynamics is given by the weakly disordered Anderson model, specifically, $h=-\Delta+\gamma V$ with $V=\{V(x)\}_{x\in\mathbb Z^D}$ an i.i.d.\ family of Gaussian random variables. It is widely expected \cite{aizenman2015random} that this model exhibits quantum diffusion in dimensions $D\geq 3$. Recently, a proof of the closely related absolute continuity of the spectrum was claimed by OpenAI \cite{OpenAI_Anderson_3D}. Quantum diffusion has been proved on perturbatively short time scales in dimensions $D\geq 2$ in \cite{Erdos_2007_1,Erdos_2007_2,Erdos_2008,BlackDroginHernandez2025,hernandez2024quantum}. 
	In Corollary~\ref{cor:anderson}, we  combine Theorem~\ref{thm:transport-stability} with the quantum diffusion result of Black--Drogin--Hern\'andez \cite{BlackDroginHernandez2025} extending  earlier work of Erd{\H o}s-Salmhofer-Yau \cite{Erdos_2007_1,Erdos_2007_2,Erdos_2008}. Roughly speaking, our first result establishes the (time-averaged) many-body diffusion hallmark:
	\begin{equation}
		\label{eq:introdiffusion}
		\frac{1}{T}\int_0^T\sum_{x\in \Lambda} |x|^2 \abs{\ac{\tau_t^\varepsilon(a_0^*)}{a_x}}^2 \dd t
		\approx T\one+\mathcal O(\varepsilon  T^{D+3}).
	\end{equation}
	for  times $\gamma^{-2}\leq T\leq \gamma^{-2-\kappa_D}$ with $\kappa_D>0$ an explicit, small, dimension-dependent constant from \cite{BlackDroginHernandez2025}, as in \eqref{eq:kappa}. 
	We refer to Corollary~\ref{cor:anderson} for the precise statement, which actually treats the upper and lower bounds slightly differently as in \cite{BlackDroginHernandez2025}. The diffusion scaling can then be observed for  times \[t\leq \min\{\gamma^{-2-\kappa_D},\varepsilon ^{-1/(D+2)}\}.\]

	This is a rare rigorous result on  diffusive behavior of a disordered interacting \textit{many-body} system. We emphasize that it heavily uses the one-body result \cite{BlackDroginHernandez2025} and that it is only a result about the specific operator-spreading diffusion diagnostic
	\[
	\sum_{x\in \Lambda} |x|^2 \abs{\ac{\tau_{t}^{\varepsilon}(a_0^*)}{a_x}}^2
	\]
	and its normalized version. In particular, we do not claim to prove the much stronger quantum diffusion statement that the dynamics is approximated by a Boltzmann equation.

	\subsection{Second Result: Fermi-Hubbard Hamiltonians at low density.}
	We asked ourselves if locality can also be used to control the effect of interactions if the small parameter $\varepsilon$ does not appear in the Hamiltonian (like the weak coupling  constant in \eqref{eq:Hresult1intro}), but rather in the \textit{initial state}. This question is inspired by recent progress in state-dependent propagation bounds for Bose-Hubbard Hamiltonians, which sensitively depend on the initial state's particle density \cite{yin2022finite,faupin2022maximal,kuwahara2024effective,lemm2025quantum,faupin2025macroscopic,LRZ2026}. To explore this question, we consider  a broad class of   Fermi-Hubbard type Hamiltonians with two spin species:
	\[
	H_{\mathrm{FH}}=T_\uparrow +T_\downarrow
	+\sum_{x\in \Lambda}U_xn_{x,\uparrow}n_{x,\downarrow},\qquad T_\sigma=\sum_{\substack{x,y\in \Lambda:\\ x\sim y}}
	J_{xy,\sigma}a_{x,\sigma}^*a_{y,\sigma}\] 
	Here $\sigma\in\{\uparrow,\downarrow\}$ and \(J_{xy,\sigma},U_x\) are not necessarily small. Instead, the relevant small parameter in this case is taken to be the spin-down density in the initial state $\rho$:
	\[\mathrm{Tr}(\rho n_{x\downarrow})\leq \zeta\ll 1.
	\]Notice that a small spin-down density should also suppress the interaction term $n_{x,\uparrow}n_{x,\downarrow}$ and this is precisely what we aim to make rigorous.
	Indeed, Theorem~\ref{thm:Hubbard} approximates every local multi-time correlation function of local spin-up observables $A_1,\ldots, A_n$ by the free counterpart evolved via the free Hamiltonian
	\[
	H_{\mathrm{f}}=T_\uparrow+T_\downarrow
	\]
	Then, Theorem~\ref{thm:Hubbard}  says, roughly speaking, that
	\begin{equation}
		\begin{aligned}
			&\Tr\!\left(\rho\prod_{j=1}^n\tauFH_{t_j}(A_j)\right)
			-\Tr\!\left(\rho\prod_{j=1}^n\tauf_{t_j}(A_j)\right)\\
			&\qquad=\mathcal O\!\left((1+T)^{D-1}\e^{-vT}
			+\zeta\,(1+T)^{D}\right)
			\label{eq:intro_Hubbard-window}
	\end{aligned}\end{equation} 
	for arbitrary (not necessarily ordered) times $t_1,\ldots,t_n\in [0,T]$ for $T > 0$ large. 
	In particular, in Corollary~\ref{cor:OTOC}, we specialize to the creation--annihilation out-of-time-ordered correlator (OTOC).  Together with an exact one-body calculation in Section~\ref{sec:free-laplacian-OTOC}, this provides an OTOC analog of the ``diagonal-heavy'' pattern of short-time two-site entanglement distribution observed numerically in Fermi--Hubbard Hamiltonians \cite[Figure 4]{bippus2026two}. We refer to the end of Section \ref{sec:free-laplacian-OTOC} for a more detailed discussion.

	\subsection{Outlook} Taking a broader perspective, we would like to emphasize that our applications, which used first-order Duhamel identities to compare with one-body reference dynamics, can be substantially refined in a rather obvious way.    In future work, we plan to combine Lieb--Robinson ideas with higher-order Duhamel expansions of the schematic form
	\begin{equation}\label{eq:intro-higher-order}
		\tau_t(A)
		=\tau_t^{\rm ref}(A)
		+\sum_{k=1}^{m}\mathcal D_k(t;A)
		+\mathcal R_{m+1}(t;A),
	\end{equation}
	where the ordered integrals \(\mathcal D_k\) contain nested local commutators.
	Locality will make the remainder \(\mathcal R_{m+1}\) uniform in system size. The goal is to obtain controlled approximations by effective nonlinear evolution equations in longer-time dynamical regimes, where comparison with a free one-body dynamics is too crude, and one should instead use higher-order effective dynamics.  We expect that locality, as captured by the LRB, can provide the mathematical foundation to such a new research program of deriving finer effective evolution equations on perturbative time scales beyond mean-field regimes or related scaling limits.
	
	\subsection{Organization of the paper.}
	In Sections~\ref{sec:transport} and \ref{sec:Hubbard} we present our two results concerning transport in weakly interacting fermions and the low-density Fermi Hubbard model, respectively. 
	The lattice estimates used in the transport application in Section \ref{sec:transport} are collected in Appendix~\ref{sec:tails}.
	
	\section{Weak interactions and stability of transport}\label{sec:transport}
	In this section, we present our first main result, Theorem~\ref{thm:transport-stability} below, and its application to many-body diffusion in Corollary~\ref{cor:anderson}. 
	\subsection{Model and transport quantity}

	Throughout this section, \(\lvert\cdot\rvert\) denotes the \(\ell^1\)-distance on \(\mathbb Z^D\).  Let \(h\) be a self-adjoint one-body operator on \(\ell^2(\mathbb Z^D)\), with matrix elements \(h_{xy}=\langle\delta_x,h\delta_y\rangle\).  We assume that every \(\delta_x\) lies in \(\operatorname{Dom}(h)\) and that, on the domain of its diagonal multiplication operator, \(h\) is the sum of that diagonal operator and the off-diagonal matrix \((h_{xy})_{x\ne y}\).  For a finite \(\Lambda\subset\mathbb Z^D\) containing the origin, \(h_\Lambda\) denotes the open restriction (meaning the restriction to $\Lambda$ with open boundary conditions) of \(h\) to \(\ell^2(\Lambda)\).
	For \(r\geq0\), write
	\begin{equation}\label{eq:Brdefn}
		B_r(x):=\{y\in\mathbb Z^D:|y-x|\leq r\}.
	\end{equation}
	
	We work on the fermionic Fock space \(\F_\Lambda=\bigoplus_{n\geq0}\bigwedge^n\ell^2(\Lambda)\).  The creation and annihilation operators \(a_x^*,a_x\) satisfy
	\begin{equation}\label{eq:CAR-spinless}
		\ac{a_x}{a_y^*}=\delta_{xy}\one,
		\qquad \ac{a_x}{a_y}=\ac{a_x^*}{a_y^*}=0,
		\qquad n_x:=a_x^*a_x.
	\end{equation}
	
	Let \(U_{xy}=U_{yx}\in\mathbb R\), \(x,y\in\mathbb Z^D\), with \(U_{xx}=0\).  For \(\varepsilon\in\mathbb R\), consider
	\begin{equation}\label{eq:transport-H}
		\begin{aligned}
			H_\varepsilon&=T+\varepsilon W,\\
			T&=\sum_{x,y\in\Lambda}h_{xy}a_x^*a_y,
			&
			W&=\sum_{x,y\in\Lambda}U_{xy}n_xn_y.
		\end{aligned}
	\end{equation}
	
	\begin{assumption}[Finite-range]
		\label{ass:transport-locality}
		There are constants \(R_*,Q_*>0\) such that
		\begin{equation}\label{eq:h-row-sum}
			h_{xy}=U_{xy}=0\quad\text{if }|x-y|>R_*,
			\qquad
			\sup_{x\in\mathbb Z^D}\sum_{\substack{y\in\mathbb Z^D\\y\ne x}}
			(|h_{xy}|+|U_{xy}|)\leq Q_*.
		\end{equation}
	\end{assumption}
	No bound is imposed on the diagonal entries \(h_{xx}\). 
	
	We write \(\tau_t^\varepsilon\) for the Heisenberg dynamics generated by \(H_\varepsilon\).
	
	\begin{definition}[Transport kernel and moments]
		For \(t\geq0\), define the finite-volume interacting moment and the whole-space one-body (``free'') moment by
		\begin{align}
			M_{\varepsilon}(t)&:=\sum_{x\in\Lambda}|x|^2\abs{\ac{\tau_t^\varepsilon(a_0^*)}{a_x}}^2\in\mathcal B(\F_\Lambda),
			\label{eq:M-Lambda}\\
			\bar M_{\varepsilon}(t)&:=S_\varepsilon(t)^{-1/2}M_{\varepsilon}(t)S_\varepsilon(t)^{-1/2}\in\mathcal B(\F_\Lambda),
			\label{eq:M-Lambda-normalized}\\
			M_{\mathrm{f}}(t)&:=\sum_{x\in\mathbb Z^D}|x|^2
			\abs{\ip{\delta_x}{\e^{\iu th}\delta_0}}^2.\label{eq:M-onebody}
		\end{align}
	\end{definition}
	Here, $\bar M_{\varepsilon}(t)$ is defined when $S_\varepsilon(t)$ defined in \eqref{eq:S_def} is invertible. It is the second moment of the positive operator-valued analog of a probability measure
	\begin{equation*}
		x\mapsto S_\varepsilon(t)^{-1/2}\abs{\ac{\tau_t^\varepsilon(a_0^*)}{a_x}}^2S_\varepsilon(t)^{-1/2}.
	\end{equation*}
	Taking the expectation of this measure in any state gives a probability measure.

	\subsection{General comparison theorem} We now state our first main result, providing a general comparison for the one-body and many-body transport kernels from \eqref{eq:M-Lambda}--\eqref{eq:M-onebody}. 
	
	\begin{theorem}[General comparison theorem]
		\label{thm:transport-stability}
		Suppose Assumption~\ref{ass:transport-locality} holds and fix \(\varepsilon_0>0\).  There are constants \(C,v,R_0>0\), depending only on \(D,R_*,Q_*,\varepsilon_0\), with the following property.  If \(B_R(0)\subset\Lambda\), \(R\geq R_0\), and \(\varepsilon \in(0,\varepsilon_0)\), then, for every \(t\geq0\),
		\begin{align}
			\norm{M_\varepsilon(t)-M_{\mathrm{f}}(t)\one}
			\leq{}&C(1+R)^{D+2}\varepsilon \,t
			\left[1+\varepsilon t(1+R)^D\right]\nonumber\\
			&+C(1+R)^{D+1}(1+t)\e^{2vt- R}.
			\label{eq:moment-full-comparison}
		\end{align}
		Suppose that the right-hand side of \eqref{eq:moment-full-comparison} is $<1/2$, then we also have the normalized analog 
		\begin{align}
			\norm{\bar M_\varepsilon(t)-M_{\mathrm{f}}(t)\one}
			\leq{}&C(1+R)^{D+2}\varepsilon \,t
			\left[1+\varepsilon t(1+R)^D\right]\nonumber\\
			&+C(1+R)^{D+1}(1+t)\e^{2vt- R}.
			\label{eq:moment-full-comparison-normalized}
		\end{align}
	\end{theorem}

	The proof, given in Section \ref{subsec:proof-compare} below, makes all locality inputs explicit.  The fermionic Lieb--Robinson estimate controls the interacting tail, while a one-particle finite-propagation estimate replaces the finite-volume free reference by the full-space object in \eqref{eq:M-onebody}. 
	
	\begin{remark}
		The fact that the underlying Lieb-Robinson bounds can be made independent of the diagonal entries $h_{xx}$ follows by a well-known interaction picture trick \cite{nachtergaele2007lieb}. The same trick works for fermions. Intuitively, it is made possible from the fact that they do not produce operator spreading. The independence of the resulting bound  \eqref{eq:moment-full-comparison} of the size of the $h_{xx}$ is useful for our next application to the Anderson model with Gaussian (hence, unbounded) disorder. We would like to further stress that since one of our major interests here is to establish the stability of delocalization, as quantified by the norms of the operators in equations \eqref{eq:introexp} and \eqref{eq:introexp_normalized}, we can ignore the dependence of the LRBs by random potential, namely the diagonal elements $h_{xx}$. On the other hand taking this into account has been obviously crucial when proving, for example, the fact that the LR velocity of the one dimensional $XX$ model with a random $Z$ field vanishes \cite{hamza2012dynamical}.   
	\end{remark}
	
	\subsection{Application to many-body diffusion}
	
	In this section, we specialize to the case where the non-interacting part is given by the weakly disordered Anderson model
	\begin{equation}\label{eq:Anderson}
		h_\gamma=-\Delta_{\mathbb Z^D}+\gamma V
		\quad\text{on }\ell^2(\mathbb Z^D),
		\qquad
		(-\Delta_{\mathbb Z^D}\psi)(x)
		:=2D\psi(x)-\sum_{y:y\sim x}\psi(y),
	\end{equation}
	where \(\{V(x)\}_{x\in\mathbb Z^D}\) are independent standard Gaussian random variables and the coupling constant $\gamma>0$ is another small parameter.  The multiplication operator \(V\) is almost surely unbounded, but it is self-adjoint on its natural domain and \(-\Delta_{\mathbb Z^D}\) is bounded.  More importantly for us, the one-particle bounds \eqref{eq:onebody-light-cone} and \eqref{eq:onebody-boundary} are uniform in every real diagonal potential.
	
	Black, Drogin, and Hern\'andez proved time-averaged delocalization and concentration estimates in \(D\geq2\) \cite[Theorems~1.1 and~5.2]{BlackDroginHernandez2025}.  We use their result (which has a different kinetic-energy convention but produces the same estimates). Since $h_\gamma$ has real matrix entries, complex conjugation shows that its transition probabilities are unchanged by $t\mapsto-t$. For every \(\delta>0\), \cite[Theorems~1.1 and~5.2]{BlackDroginHernandez2025} establish that there are \(c_\delta,C_\delta>0\) such that, for each fixed \(T\) in the range \eqref{eq:BDH-window}, with probability at least \(1-C_\delta\gamma^{1000}\),
	\begin{equation}\label{eq:BDH}
		\frac1T\int_0^T\sum_{x\in\mathbb Z^D}|x|^2
		\abs{\ip{\delta_x}{\e^{\iu t h_\gamma}\delta_0}}^2\,\dd t
		\geq c_\delta\gamma^{-2}T
	\end{equation}
	and
	\begin{equation}\label{eq:BDH-escape}
		\frac1T\int_0^T
		\sum_{|x|\geq C_\delta\gamma^{-1}\sqrt T}
		\abs{\ip{\delta_x}{\e^{\iu t h_\gamma}\delta_0}}^2\,\dd t
		\leq\delta,
	\end{equation}
	whenever, for sufficiently small \(\gamma\),
	\begin{equation}\label{eq:BDH-window}
		\max\{C_\delta,\gamma^{-2}\}
		\leq T\leq\gamma^{-2-\kappa_D+\delta}.
	\end{equation}
	Here, the explicit dimension-dependent choices of $\kappa_D>0$ are
	\begin{equation}\label{eq:kappa}
		\kappa_D=
		\begin{cases}
			2/13,&D\in\{2,4,5,6\},\\
			2/9,&D=3,\\
			(2D-12)/(3D-12),&7\leq D\leq12,\\
			1/2,&D>12.
		\end{cases}
	\end{equation}
	
	In the general setup above, take \(h=h_\gamma\), its open restriction \(h_\Lambda=h_{\gamma,\Lambda}\), and \(U_{xy}=1\) for nearest neighbors and zero otherwise.    Assumption~\ref{ass:transport-locality} holds uniformly with \(R_*=1\) and \(Q_*=4D\).
	
	\begin{corollary}[Finite-volume short-time diffusion transfer]
		\label{cor:anderson}
		Fix \(0<\delta<\kappa_D\) and \(\theta\in(0,1)\).  There are constants \(a,C,c_{\delta,\theta}>0\) such that the following holds for all sufficiently small \(\gamma\).  Let \(T\) satisfy \eqref{eq:BDH-window} and
		\begin{equation}\label{eq:epsilon-threshold}
			T\leq c_{\delta,\theta} \varepsilon^{-1/(D+2)} 
		\end{equation}
		and  let \(\Lambda\) contain \(B_{a(1+T)}(0)\).
		Then, with probability at least \(1-C_\delta\gamma^{1000}\),
		\begin{equation}\label{eq:interacting-diffusion}
			\frac1T\int_0^T M_\varepsilon(t)\,\dd t
			\geq\left((1-\theta)c_\delta\gamma^{-2}T-\frac CT\right)\one,
		\end{equation}
		and
		\begin{equation}
			\frac1T\int_0^T
			\sum_{\substack{x\in\Lambda\\
					|x|\geq C_\delta\gamma^{-1}\sqrt T}}
			\abs{\ac{\tau_t^\varepsilon(a_0^*)}{a_x}}^2\,\dd t
			\leq2\delta\one.\label{eq:interacting-escape}
		\end{equation}
		Both inequalities are in the operator order sense on $\mathcal B(\F_\Lambda)$. At the upper end of \eqref{eq:BDH-window}, the sufficient condition \eqref{eq:epsilon-threshold} has the scaling
		\begin{equation}
			\varepsilon
			\lesssim \gamma^{(2+\kappa_D-\delta)(D+2)}.
		\end{equation}
	\end{corollary}

	We remark that the term $-\frac{C}{T}$ in \eqref{eq:interacting-diffusion} is unimportant in the regime $T\geq1$ that we are interested in. It arises from the prefactor $\frac{1}{T}$ in the time average. If one would like a bounded estimate for small $T$, then it is possible to replace $\frac{1}{T}$ by a constant using the standard short-time improvement of the LRB \eqref{eq:fermion-LR} in which $e^{vt}-1$ takes the place of $e^{vt}$.

	\subsection{Proof of the comparison theorem} \label{subsec:proof-compare}
	In this section, we provide the proof of Theorem~\ref{thm:transport-stability}. In the following, for a finite \(\Omega\subset\mathbb Z^D\), let \(H_{\varepsilon,\Omega}\), \(W_\Omega\), and \(\tau_t^{\varepsilon,\Omega}\) denote the open restrictions of the objects in \eqref{eq:transport-H} to \(\Omega\).

	\begin{proof}[Proof of Theorem~\ref{thm:transport-stability}]
		In general, the proof follows the procedure (I)--(III) outlined in the introduction: (I), localize the many-body dynamics to \(B_R(0)\), (II), compare the two dynamics in that ball, and, (III), replace the restricted one-body dynamics by the whole-space one. The  difference is that we do not work with the Heisenberg dynamics of a local bounded observable since the second moment lives on the full space and has a growing weight.  
		
		\smallskip\noindent\emph{Step (I): Initial localization.}
		We first state the relevant fermionic LRB.
		\begin{lemma}\label{lm:fermion-LR}
			There exist constants \(C_{\rm LR},v>0\) such that, for every finite \(\Omega\),
			\begin{equation}\label{eq:fermion-LR}
				\norm{\ac{\tau_t^{\varepsilon,\Omega}(a_y^*)}{a_x}}+	\norm{\ac{\tau_t^{\varepsilon,\Omega}(a_y)}{a_x}}
				\leq C_{\rm LR}\e^{vt-4|x-y|},
				\qquad t\geq0.
			\end{equation}
		\end{lemma}
		
		The spatial decay rate $4$ is chosen only for simplicity to avoid an extra parameter; any other decay rate is also possible and will affect the velocity $v$.
		
		\begin{proof}[Proof of Lemma \ref{lm:fermion-LR}]
			This is essentially the standard fermionic LRB from \cite[Theorem~3.1(ii)]{NachtergaeleSimsYoung2018Fermions}, but we briefly explain how the on-site dependence can be removed via a fermionic version of the usual interaction picture trick going back to \cite{nachtergaele2007lieb}. Let
			\begin{equation}
				H_{\mathrm{diag},\Lambda}:=\sum_{x\in\Lambda}h_{xx}n_x
			\end{equation}
			and pass to its interaction picture.  For each unordered pair \(\{x,y\}\), the resulting even interaction is
			\begin{align}
				\widetilde\Phi_{\varepsilon,s}(\{x,y\})
				={}&h_{xy}\e^{\iu s(h_{xx}-h_{yy})}a_x^*a_y
				+\overline{h_{xy}}\e^{-\iu s(h_{xx}-h_{yy})}a_y^*a_x\nonumber\\
				&+2\varepsilon U_{xy}n_xn_y.
				\label{eq:transport-IP-interaction}
			\end{align}
			Its support and range are unchanged, and
			\begin{equation}
				\norm{\widetilde\Phi_{\varepsilon,s}(\{x,y\})}
				\leq2|h_{xy}|+2\varepsilon_0|U_{xy}|.
			\end{equation}
			Finite range and \eqref{eq:h-row-sum} therefore bound every exponentially weighted interaction norm required by the fermionic Lieb--Robinson theorem, uniformly in time and in the diagonal entries.  Applying \cite[Theorem~3.1(ii)]{NachtergaeleSimsYoung2018Fermions} with spatial decay rate equal to 4 proves the lemma.
		\end{proof}
		
		Set \(B:=B_R(0)\). We recall how combining Duhamel's formula with the Lieb--Robinson theorem gives a local approximation of the Heisenberg dynamics:
		\begin{equation}\label{eq:manybody-localization}
			\norm{\tau_t^{\varepsilon,\Lambda}(a_0^*)
				-\tau_t^{\varepsilon,B}(a_0^*)}
			\leq C t(1+R)^{D-1}\e^{vt-4R}.
		\end{equation}
		To see this, recall Definition \eqref{eq:Hresult1intro} of our Hamiltonian, let \(H_{\varepsilon,\Lambda}^{\rm dec}
		:=H_{\varepsilon,B}+H_{\varepsilon,\Lambda\setminus B}\) and write \(H_{\partial B}:=H_{\varepsilon,\Lambda}
		-H_{\varepsilon,\Lambda}^{\rm dec}
		\)
		Note that $H_{\partial B}$ consists exactly of terms intersecting both $B$ and $B^c$. Accordingly, we may write
		\[
		H_{\partial B}=\sum_{\substack{Z:Z\cap B\neq\emptyset\\ Z\cap B^C\neq\emptyset }} \Phi_\varepsilon (Z)
		\]
		Duhamel's formula then  gives
		\begin{equation}
			\norm{\tau_t^{\varepsilon,\Lambda}(a_0^*)
				-\tau_t^{\varepsilon,B}(a_0^*)} \leq \int_0^t\sum_{\substack{Z:Z\cap B\neq\emptyset\\ Z\cap B^C\neq\emptyset }} 
			\norm{[\Phi_\varepsilon(Z),
				\tau_{t-s}^{\varepsilon,B}(a_0^*)]}\,\dd s.
		\end{equation}
		Every term \(\Phi_\varepsilon(Z)\) in this sum is even.  Expressing commutators as anti-commutators via $[AB,C]=A\{B,C\}-\{A,C\} B$ and using $\|a_x\|=1$, we can invoke Lemma \ref{lm:fermion-LR} to bound 
		\[
		\norm{[\Phi_\varepsilon(Z),
			\tau_{t-s}^{\varepsilon,B}(a_0^*)]}
		\leq C\norm{\Phi_\varepsilon(Z)}
		\e^{vt-4\dist(0,Z)}
		\]
		The assumption \eqref{eq:h-row-sum} gives total boundary strength at most \(C(1+R)^{D-1}\), while \(\dist(0,Z)\geq R-R_*\).  Integration proves \eqref{eq:manybody-localization}; the harmless shift by \(R_*\) is absorbed into \(C\).
		
		We continue with Step (II). Define the restricted interacting moment
		\begin{equation}
			M_{\varepsilon,B}(t):=
			\sum_{x\in B}|x|^2
			\abs{\ac{\tau_t^{\varepsilon,B}(a_0^*)}{a_x}}^2.
		\end{equation}
		We regard operators supported in $B$ as operators on $\F_\Lambda$. By the triangle inequality and 
		\begin{align} \nonumber
			\norm{A^*A-B^*B} = \norm{A^*A - A^*B + A^*B -B^*B} \leq(\norm A+\norm B)\norm{A-B}
		\end{align}
		it holds:
		\[
		\begin{aligned}
			&\norm{M_\varepsilon(t)-M_{\varepsilon,B}(t)}\\
			& \leq \sum_{x\in B} |x|^2
			\norm{\abs{\ac{\tau_t^{\varepsilon,\Lambda}(a_0^*)}{a_x}}^2-\abs{\ac{\tau_t^{\varepsilon,B}(a_0^*)}{a_x}}^2}\\
			&+\sum_{x\in B^c} |x|^2 \norm{\ac{\tau_t^{\varepsilon,\Lambda}(a_0^*)}{a_x}}^2\\
			& \leq  8\sum_{x\in B} |x|^2
			\norm{\tau_t^{\varepsilon,\Lambda}(a_0^*)
				-\tau_t^{\varepsilon,B}(a_0^*)}
			+\sum_{x\in B^c} |x|^2 \norm{\ac{\tau_t^{\varepsilon,\Lambda}(a_0^*)}{a_x}}^2
		\end{aligned}
		\]
		Now we apply \eqref{eq:manybody-localization} and \eqref{eq:fermion-LR} inside, respectively, outside $B$:
		\[
		\begin{aligned}
			\norm{M_\varepsilon(t)-M_{\varepsilon,B}(t)}
			\leq  C\sum_{x\in B} |x|^2
			Ct (1+R)^{D-1}\e^{vt-4R}
			+\sum_{x\in B^c} |x|^2 C_{\mathrm{LR}} e^{2vt-8|x|}
		\end{aligned}
		\]
		Estimating the lattice sums by Lemma~\ref{lem:lattice-tails}, we obtain
		\begin{equation}\label{eq:manybody-moment-localization}
			\norm{M_\varepsilon(t)-M_{\varepsilon,B}(t)}
			\leq C(1+R)^{2D+2}(1+t)\e^{2vt-4R}
			\leq C(1+t)\e^{2vt-R}
		\end{equation}
		up to changing the constant $C$.

		\smallskip\noindent\emph{Step (II): Local comparison.}
		Let
		\begin{equation}
			\begin{aligned}
				M_{{\rm f},B}(t)&:=\sum_{x\in B}|x|^2
				\abs{\ip{\delta_x}{\e^{\iu th_B}\delta_0}}^2,\\
				M_{{\rm f},B}(t)\one&=\sum_{x\in B}|x|^2
				\abs{\ac{\tau_t^{0,B}(a_0^*)}{a_x}}^2.
			\end{aligned}
		\end{equation}
		In this step, we aim to estimate $\norm{M_{\varepsilon,B}(t)-M_{{\rm f},B}(t)\one}$.
		The relevant Duhamel identity is
		\begin{equation}\label{eq:kernel-Duhamel}
			\tau_t^{\varepsilon,B}(a_0^*)-\tau_t^{0,B}(a_0^*)
			=\iu\varepsilon\int_0^t
			\tau_s^{\varepsilon,B}\!\left(
			[W_B,\tau_{t-s}^{0,B}(a_0^*)]\right)\dd s.
		\end{equation}
		A naive estimate on this is $\varepsilon  t$ times $\|W_B\| \approx R^{D}$. Multiplying by $|x|^2$ and summing over $B$, this would lead to an error $\varepsilon t R^{2D+2}$. Intuitively, this should be improvable by exploiting the fact that $\tau_{t-s}^{0,B}(a_0^*)$ does not explore the full perturbation $W_B$ at once. This can be made precise using the fact that this dynamics is in fact non-interacting, as we explain now.

		For \(f\in\ell^2(B)\), write \(a^*(f):=\sum_{z\in B}f(z)a_z^*\).  The CAR and the ordered-pair convention in \eqref{eq:transport-H} give
		\begin{align}
			[W_B,a^*(f)]
			&=\sum_{z\in B}f(z)a_z^*K_z,
			&K_z&:=2\sum_{y\in B}U_{zy}n_y,
			&\norm{K_z}&\leq2Q_* . \label{eq:ball-W-commutator}
		\end{align}
		Consequently,
		\begin{equation}\label{eq:ball-W-l1}
			\norm{[W_B,a^*(f)]}
			\leq2Q_*\norm f_{\ell^1(B)}
			\leq2Q_*|B|^{1/2}\norm f_{\ell^2(B)}.
		\end{equation}
		Since \(\tau_r^{0,B}(a_0^*)=a^*(\e^{\iu rh_B}\delta_0)\), the one-body vector $f$ appearing in \eqref{eq:ball-W-l1} has \(\ell^2\)-norm one.  Thus \eqref{eq:ball-W-l1} and \(|B|\leq C(1+R)^D\) yield
		\begin{equation}\label{eq:ball-field-Duhamel}
			d_R(t):=
			\norm{\tau_t^{\varepsilon,B}(a_0^*)-
				\tau_t^{0,B}(a_0^*)}
			\leq C\varepsilon t(1+R)^{D/2}.
		\end{equation}
		
		We set
		\begin{equation}
			\begin{aligned}
				p_x(t)&:=\ac{\tau_t^{\varepsilon,B}(a_0^*)}{a_x}\in\mathcal B(\F_\Lambda),\\
				q_x(t)&:=\abs{\ip{\delta_x}{\e^{\iu th_B}\delta_0}}.
			\end{aligned}
		\end{equation}
		The free CAR identity gives \(\{\tau_t^{0,B}(a_0^*),a_x\}
		=\langle\delta_x,\e^{\iu th_B}\delta_0\rangle\one\), and hence
		\begin{equation}\label{eq:kernel-error}
			\norm{p_x(t)^*p_x(t)-q_x(t)^2\one}
			\leq4d_R(t)q_x(t)+4d_R(t)^2.
		\end{equation}
		
		By Cauchy--Schwarz and one-body unitarity,
		\begin{align}
			\sum_{x\in B}|x|^2q_x(t)
			&\leq
			\left(\sum_{x\in B}|x|^4\right)^{1/2}
			\left(\sum_{x\in B}q_x(t)^2\right)^{1/2}
			\leq C(1+R)^{D/2+2},\\
			\sum_{x\in B}q_x(t)&\leq C(1+R)^{D/2}.
		\end{align}
		Together with \eqref{eq:ball-field-Duhamel}, this proves
		\begin{equation}\label{eq:restricted-moment-comparison}
			\norm{M_{\varepsilon,B}(t)-M_{{\rm f},B}(t)\one}
			\leq C\varepsilon t(1+R)^{D+2}
			\left[1+\varepsilon t(1+R)^D\right].
		\end{equation}
		This is a substantial improvement over the naive estimate $\varepsilon t R^{2D+2}$.
		We will also use the corresponding unweighted estimate
		\begin{equation}\label{eq:restricted-mass-comparison}
			\sum_{x\in B}\norm{p_x(t)^*p_x(t)-q_x(t)^2\one}
			\leq C\varepsilon t(1+R)^D
			\left[1+\varepsilon t(1+R)^D\right].
		\end{equation}
		
		\smallskip\noindent\emph{Step (III): Extension of the one-body dynamics.}
		Set
		\begin{equation}
			v_0 := 1 + \sup_x\sum_{y\ne x}|h_{xy}|\e^{4|x-y|}
			\leq 1+ \e^{4R_*}Q_*
			\label{eq:Jmu-v0}
		\end{equation}
		A standard one-body propagation bound, see e.g. \cite[Appendix~B, Equations~(B.1)--(B.4)]{aizenman2012absolutely} then gives
		\begin{equation}\label{eq:onebody-light-cone}
			\abs{\ip{\delta_x}{\e^{\iu t h_\Omega}\delta_y}}
			\leq\e^{v_0t-4|x-y|},
			\qquad t\geq0.
		\end{equation}
		In particular, the constants are independent of the diagonal part $h_{xx}$. (The estimate \eqref{eq:onebody-light-cone} can be proved, e.g., by passing to the interaction picture of the diagonal part of \(h\) and performing a weighted Dyson expansion.)

		If \(B_b(y)\subset\Omega\), delete the hopping terms crossing the boundary of the common ball \(B_b(y)\), both for \(h\) and for \(h_\Omega\).  The two decoupled evolutions agree on \(\delta_y\).  Applying Duhamel's formula to each comparison, then the triangle inequality and \eqref{eq:onebody-light-cone}, yields
		\begin{equation}\label{eq:onebody-boundary}
			\norm{\e^{\iu th}\delta_y-\e^{\iu th_\Omega}\delta_y}
			\leq C(1+b)^{D-1}\e^{-4b}
			\frac{\e^{v_0t}-1}{v_0},
			\qquad t\geq0.
		\end{equation}
		
		Extend \(u_B(t):=\e^{\iu th_B}\delta_0\) by zero to \(\ell^2(\mathbb Z^D)\), and put \(u(t):=\e^{\iu th}\delta_0\).  Applying \eqref{eq:onebody-boundary} with \(b=R\), and using the multiplication operator \(X_R(x)=|x|^2\mathbf1_{\{|x|\leq R\}}\), gives
		\begin{equation}
			\abs{\ip{u_B(t)}{X_Ru_B(t)}-\ip{u(t)}{X_Ru(t)}}
			\leq C R^2(1+R)^{D-1}\e^{-4R}
			\frac{\e^{v_0t}-1}{v_0}.
		\end{equation}
		The one-body light cone and Lemma~\ref{lem:lattice-tails} also give
		\begin{equation}
			\sum_{|x|>R}|x|^2
			\abs{\ip{\delta_x}{\e^{\iu th}\delta_0}}^2
			\leq C(1+R)^{D+1}\e^{2v_0t-8R}.
		\end{equation}
		After enlarging \(v\) so that \(v\geq v_0\), these two estimates imply
		\begin{equation}\label{eq:onebody-moment-extension}
			\abs{M_{{\rm f},B}(t)-M_{\mathrm f}(t)}
			\leq C(1+R)^{D+1}(1+t)\e^{2vt-R}.
		\end{equation}
		Combining \eqref{eq:manybody-moment-localization}, \eqref{eq:restricted-moment-comparison}, and \eqref{eq:onebody-moment-extension} proves \eqref{eq:moment-full-comparison}.
		
		To prove \eqref{eq:moment-full-comparison-normalized}, repeat the localization argument without the weight $|x|^2$ and use \eqref{eq:restricted-mass-comparison}. Since the restricted free mass is one, this gives
		\begin{align*}
			\norm{S_\varepsilon(t)-\one}
			\leq{}&C\varepsilon t(1+R)^D\left[1+\varepsilon t(1+R)^D\right]\\
			&+C(1+R)^{D-1}(1+t)\e^{2vt-R}.
		\end{align*}
		By the smallness assumption in the theorem, this is $<1/2$, after choosing its constant $C$ large enough. Thus $\norm{S_\varepsilon(t)^{-1}}\leq2$ and $\norm{S_\varepsilon(t)^{-1}-\one}\leq2\norm{S_\varepsilon(t)-\one}$. Using $0\leq M_{{\rm f},B}(t)\leq R^2$, we obtain
		\begin{align*}
			\norm{\bar M_\varepsilon(t)-M_{\mathrm f}(t)\one}
			\leq{}&2\norm{M_\varepsilon(t)-M_{{\rm f},B}(t)\one}\\
			&+2R^2\norm{S_\varepsilon(t)-\one}
			+\abs{M_{{\rm f},B}(t)-M_{\mathrm f}(t)}.
		\end{align*}
		Combining this with \eqref{eq:manybody-moment-localization}, \eqref{eq:restricted-moment-comparison}, and \eqref{eq:onebody-moment-extension} proves \eqref{eq:moment-full-comparison-normalized}.
	\end{proof}

	\begin{remark}[Perturbative time scale]\label{rem:transport-time}
		Take \(R=c(1+t)\) with \(c>2v\).  In the perturbative regime \(\varepsilon t(1+t)^D\leq1\), the perturbative term in Theorem~\ref{thm:transport-stability} is bounded by
		\begin{equation}
			C\varepsilon (1+t)^{D+3}.
		\end{equation}
		The displayed power is not asserted to be optimal.  Its significance is that the estimate is uniform in volume and supplies an explicit interval on which a free lower bound dominates the many-body correction.
	\end{remark}

	\subsection{Proof of the Anderson application} In this section, we provide the proof of Corollary~\ref{cor:anderson}. 
	The idea is that choosing $R$ as a large constant multiple of \(T\) makes all boundary and light-cone tails exponentially small, after which \eqref{eq:BDH}--\eqref{eq:BDH-escape} apply directly.

	\begin{proof}[Proof of Corollary~\ref{cor:anderson}]
		Fix \(\varepsilon_0=1\) and set \(p:=D+2\).
		By choosing \(c_{\delta,\theta}\) sufficiently large, we may assume that \(c_{\delta,\theta}\leq1\).  Since \(T\geq\gamma^{-2}\geq1\), \eqref{eq:epsilon-threshold} gives
		\[
		\varepsilon\leq c_{\delta,\theta}^{p}T^{-p}\leq1.
		\]
		Let \(v\) be the velocity in Theorem~\ref{thm:transport-stability}; as in its proof, it also dominates the one-particle velocity \(v_0\).
		Choose the constants in the corollary so that
		\begin{equation}
			\begin{gathered}
				R:=c_0(1+T),
				\qquad c_0\geq\max\{C_\delta,R_0,2v+1\},\qquad 
				a\geq c_0.
			\end{gathered}
		\end{equation}
		After decreasing \(c_{\delta,\theta}\) once more, the same hypothesis also gives, uniformly for \(0\leq t\leq T\),
		\begin{equation}\label{eq:Anderson-ball-smallness}
			\varepsilon t(1+R)^D
			\leq C\varepsilon T(1+T)^D
			\leq Cc_{\delta,\theta}^{p}T^{-1}\leq1.
		\end{equation}
		For \(0\leq t\leq T\), the one-particle boundary estimate \eqref{eq:onebody-boundary}, with \(\Omega=B=B_R(0)\) and \(b=R\), gives the bound
		\begin{equation}\label{eq:Anderson-beta}
			\norm{\e^{\iu t h_{\gamma,B}}\delta_0
				-\e^{\iu t h_\gamma}\delta_0}
			\leq C(1+T)^{D-1}\e^{-cT}.
		\end{equation}
		Its constants are independent of the disorder realization.
		
		Apply Theorem~\ref{thm:transport-stability} with the fixed radius \(R\); its volume condition follows from \(B_{a(1+T)}(0)\subset\Lambda\) and \(a\geq c_0\).  We obtain, uniformly for \(0\leq t\leq T\),
		\begin{align}
			M_\varepsilon(t)
			&\geq\left(\sum_{x\in\mathbb Z^D}|x|^2
			\abs{\ip{\delta_x}{\e^{\iu t h_\gamma}\delta_0}}^2
			-C\varepsilon (1+T)^{D+3}\right)\one\nonumber\\
			&\quad-C(1+T)^{D+1}\e^{-cT}\one.\label{eq:Anderson-lower-proof}
		\end{align}
		Averaging \eqref{eq:Anderson-lower-proof} and using \eqref{eq:BDH} therefore yields
		\begin{equation}\label{eq:Anderson-lower-average}
			\frac1T\int_0^T M_\varepsilon(t)\,\dd t
			\geq\left(c_\delta\gamma^{-2}T
			-C\varepsilon (1+T)^{D+3}-\frac CT\right)\one.
		\end{equation}
		
		For the concentration estimate, put \(r_T=C_\delta\gamma^{-1}\sqrt T\).  Then \(r_T\leq R\), because \(T\geq\gamma^{-2}\) and \(c_0\geq C_\delta\).  We use the same three steps (I)-(III) as in the theorem.  First, \eqref{eq:manybody-localization} replaces the dynamics on \(\Lambda\) by the dynamics on \(B\) for \(r_T\leq|x|\leq R\), while \eqref{eq:fermion-LR} controls the sum outside \(R\).  Second, \eqref{eq:restricted-mass-comparison} and \eqref{eq:Anderson-ball-smallness} replace the interacting dynamics on \(B\) by the free one there at a cost \(C\varepsilon(1+T)^{D+1}\).  Third, applying \eqref{eq:Anderson-beta} to the projection onto \(\{|x|\geq r_T\}\) replaces the free dynamics on \(B\) by the full-space one-body dynamics at a cost at most $C(1+T)^{D-1}\e^{-cT}$. Hence
		\begin{align}
			\sum_{\substack{x\in\Lambda\\|x|\geq r_T}}
			\abs{\ac{\tau_t^\varepsilon(a_0^*)}{a_x}}^2
			&\leq\left(\sum_{\substack{x\in\mathbb Z^D\\|x|\geq r_T}}
			\abs{\ip{\delta_x}{\e^{\iu t h_\gamma}\delta_0}}^2\right)\one\nonumber\\
			&\quad+\left(C\varepsilon (1+T)^{D+1}
			+C(1+T)^D\e^{-cT}\right)\one.\label{eq:Anderson-concentration-proof}
		\end{align}
		Averaging and using \eqref{eq:BDH-escape} gives its right-hand side with the first term replaced by $\delta\one$.
		
		It remains only to absorb the errors.  Since \(T\geq1\), \eqref{eq:epsilon-threshold} gives
		\begin{align}
			C\varepsilon(1+T)^{D+3}
			&\leq C2^{D+3}c_{\delta,\theta}^{p}T
			\leq\theta c_\delta\gamma^{-2}T,\\
			C\varepsilon(1+T)^{D+1}
			&\leq C2^{D+1}c_{\delta,\theta}^{p}T^{-1}
			\leq\frac{\delta}{2},
		\end{align}
		by choosing \(c_{\delta,\theta}\) sufficiently small; in the first line we also used \(\gamma^{-2}\geq1\).  The exponential errors are at most \(C/T\) in \eqref{eq:Anderson-lower-average} and at most \(\delta/2\) in \eqref{eq:Anderson-concentration-proof} for sufficiently small \(\gamma\).
		This proves \eqref{eq:interacting-diffusion} and \eqref{eq:interacting-escape} on the Black--Drogin--Hern\'andez event.
	\end{proof}

	\section{The low-density Fermi--Hubbard model}\label{sec:Hubbard}
	In this section, we present our second main result, Theorem~\ref{thm:Hubbard} below. The key difference with the other main result presented in this paper is that now the interaction strength itself is not small, but just the density of spin-down particles is low. That is, the small parameter occurs in the initial state instead of the Hamiltonian. 
	
	In Sections \ref{subsec:OTOC}--\ref{sec:free-laplacian-OTOC}, we apply the general result from Theorem~\ref{thm:Hubbard} to the out-of-time-ordered correlator (OTOC). 
	\subsection{Model and main result}
	We start by introducing the model under consideration and some notation. Throughout Section \ref{sec:Hubbard}, \(\Lambda\subset\mathbb Z^D\) is finite and \(\dist(x,y):=|x-y|_1\) is the ambient lattice distance.  For \(X\subset\Lambda\) and \(R\geq0\), we write
	\begin{equation}\label{eq:neighborhood}
		\begin{aligned}
			X[R]:=\{x\in\Lambda:\dist(x,X)\leq R\}\,, \qquad 
			\partial X[R]:=X[R+1]\setminus X[R].
		\end{aligned}
	\end{equation}
	and denote the local observable algebra supported in \(X\) by \(\A_X\).
	We introduce a geometric constant \(\nu_D\geq1\), independent of \(\Lambda\), such that for every \(X\subset\Lambda\) and \(R\geq1\),
	\begin{equation}\label{eq:poly-growth}
		|X[R]|\leq \nu_D |X|(1+R)^D,
		\qquad
		|\partial X[R]|\leq \nu_D |X|(1+R)^{D-1}.
	\end{equation}

	Next, we introduce the spinful one-particle space
	\begin{equation}
		\mathfrak h_\Lambda=\ell^2(\Lambda)\otimes\mathbb C^2,
		\qquad
		\F_\Lambda=\bigoplus_{n\geq0}\bigwedge^n\mathfrak h_\Lambda.
	\end{equation}
	The spinful creation and annihilation operators obey
	\begin{equation}\label{eq:CAR-spinful}
		\ac{a_{x,\sigma}}{a_{y,\sigma'}^*}
		=\delta_{xy}\delta_{\sigma\sigma'}\one,
		\qquad
		\ac{a_{x,\sigma}}{a_{y,\sigma'}}
		=\ac{a_{x,\sigma}^*}{a_{y,\sigma'}^*}=0,
	\end{equation}
	and \(n_{x,\sigma}:=a_{x,\sigma}^*a_{x,\sigma}\).  For \(\sigma\in\{\uparrow,\downarrow\}\), fix coefficients \(J_{xy,\sigma}\), \(x,y\in\mathbb Z^D\), such that
	\begin{equation} \label{eq:finiterange}
		J_{yx,\sigma}=\overline{J_{xy,\sigma}},
		\qquad
		J_{xy,\sigma}=0\quad\text{if }\dist(x,y)>R_*
	\end{equation}
	for a fixed radius \(R_*<\infty\).  Their open restriction to \(\Lambda\) defines
	\begin{equation}\label{eq:hopping}
		T_\sigma=\sum_{x,y\in\Lambda}J_{xy,\sigma}
		a_{x,\sigma}^*a_{y,\sigma},
	\end{equation}
	and we assume
	\begin{equation}\label{eq:Qstar-Hubbard}
		Q_*:=\max_\sigma\sup_{x\in\mathbb Z^D}
		\sum_{y\in\mathbb Z^D}|J_{xy,\sigma}|<\infty.
	\end{equation}
	For real numbers \(U_x\), $x \in \Lambda$, we define
	\begin{align}
		H_{\mathrm{FH}}&:=T_\uparrow+T_\downarrow
		+\sum_{x\in\Lambda}U_xn_{x,\uparrow}n_{x,\downarrow},
		\label{eq:HFH}\\
		H_{\mathrm f}&:=T_\uparrow+T_\downarrow
		\label{eq:Hfree}
	\end{align}
	and denote their dynamics by \(\tauFH_t\) and \(\tauf_t\), respectively.
	
	Finally, for \(Y\subset\Lambda\), let
	\begin{equation}
		N_{Y,\downarrow}:=\sum_{x\in Y}n_{x,\downarrow}.
	\end{equation}
	
	As mentioned above, the key assumption is that the density of spin-down particles is small, as quantified by the parameter $\zeta \in [0,1]$ below.
	\begin{assumption}[Low spin-down density]\label{ass:low-density}
		The density matrix \(\rho\) is normalized and there is \(\zeta\in[0,1]\) such that for every \(x\in\Lambda\) and \(R\geq1\),
		\begin{equation}\label{eq:low-density}
			\Tr(\rho N_{\{x\}[R],\downarrow})
			\leq\zeta(1+R)^D.
		\end{equation}
	\end{assumption}
	
	In the following, we restrict our attention to observables satisfying
	\begin{equation}\label{eq:down-preserving}
		[A,N_{X,\downarrow}]=0, 
	\end{equation}
	which includes all pure spin-up observables.  We stress that assuming commutation with \emph{total} particle number alone would not suffice for our argument, since a spin-flip operator does not preserve the spin-down vacuum sector and we would thus lose the smallness effect captured by $\zeta$.
	
	As a final preparation, for a collection of finite sets \(X_1,\ldots,X_n\), put
	\begin{equation}
		S_X:=\sum_{j=1}^n|X_j|.
	\end{equation}
	
	We can now state our second main result; its proof is given in Section \ref{subsec:proofs}.
	\begin{theorem}[Low-density Hubbard approximation on polynomial time scales]
		\label{thm:Hubbard}
		Suppose that Assumption~\ref{ass:low-density}, the finite-range hopping condition \eqref{eq:finiterange}, and \eqref{eq:Qstar-Hubbard} hold.  For a collection $X_1, ... , X_n$ of finite sets, let \(A_j\in\A_{X_j}\) satisfy \eqref{eq:down-preserving}, and let \(t_j\geq0\), \(j=1,\ldots,n\).  
		
		Then, there are constants \(C,v>0\), depending only on \(D,R_*\), and \(Q_*\), and, in particular, independent of \(\Lambda\), \(U_x\), \(\rho\), and \(R\), such that for every \(R\geq1\),
		\begin{align}
			&\abs{\Tr\!\left(\rho\prod_{j=1}^n\tauFH_{t_j}(A_j)\right)
				-\Tr\!\left(\rho\prod_{j=1}^n\tauf_{t_j}(A_j)\right)}
			\label{eq:Hubbard-main}\\
			&\quad\leq C\prod_{j=1}^n\norm{A_j}
			\left[
			Q_*(1+R)^{D-1}\e^{- R}
			\sum_{j=1}^n|X_j|\e^{vt_j}
			+\zeta S_X(1+R)^D
			\right].\nonumber
		\end{align}
		The order in each product is the displayed order and is kept fixed.
		
		If the number and supports of the observables are fixed and \(0\leq t_j\leq T\), then choosing \(R=1+2vT\) gives
		\begin{align}
			&\abs{\Tr\!\left(\rho\prod_{j=1}^n\tauFH_{t_j}(A_j)\right)
				-\Tr\!\left(\rho\prod_{j=1}^n\tauf_{t_j}(A_j)\right)}
			\label{eq:Hubbard-window}\\
			&\quad\leq C\prod_{j=1}^n\norm{A_j}
			\left((1+T)^{D-1}\e^{-vT}
			+\zeta\,(1+T)^{D}\right).\nonumber
		\end{align}
		Here \(C\) may depend on the fixed number and supports of the observables.
		In particular, for \(T=\zeta^{-\alpha}\) with \(0<\alpha<1/D\), the difference in \eqref{eq:Hubbard-window} tends to zero as \(\zeta\downarrow0\).
	\end{theorem}
	
	\begin{remark}
		We stress that the constants in \eqref{eq:Hubbard-main} are independent of the magnitude of the Hubbard interaction.  The approximation is therefore perturbative in the state, not in \(U_x\).  This is the main distinction from Theorem~\ref{thm:transport-stability}.
	\end{remark}

	\subsection{Proof of Theorem \ref{thm:Hubbard}} \label{subsec:proofs}
	The proof strategy for Theorem \ref{thm:Hubbard} follows the procedure (I)-(II) described in the introduction. However, as already stressed above, Step (II) is different here because the small parameter is not in the Hamiltonian, but rather in  (the spin-down density of)  the initial state. In particular, the result is state dependent and so in the proof we have to avoid operator norms. Instead, we exploit the small density by using Markov's inequality to control the approximation in Duhamel's formula. Step (III) is only implemented in the application to OTOCs after Corollary~\ref{cor:OTOC}, because the free dynamics is particularly nicely expressible when it acts on single creation and annihilation operators.

	\smallskip\noindent\emph{Step (I): Initial localization.} We begin with the initial localization step. 
	For \(Y\subset\Lambda\), define the restricted Hamiltonians by restricting every sum in \eqref{eq:HFH} and \eqref{eq:Hfree} to sites and hopping pairs contained in \(Y\):
	\begin{align}
		H_{\mathrm{FH},Y}
		&:=\sum_{\sigma}\sum_{x,y\in Y}J_{xy,\sigma}
		a_{x,\sigma}^*a_{y,\sigma}
		+\sum_{x\in Y}U_xn_{x,\uparrow}n_{x,\downarrow},
		\label{eq:restricted-HFH}\\
		H_{{\rm f},Y}
		&:=\sum_{\sigma}\sum_{x,y\in Y}J_{xy,\sigma}
		a_{x,\sigma}^*a_{y,\sigma}.\label{eq:restricted-Hfree}
	\end{align}
	We write
	\begin{equation}
		\tau_t^{\mathrm{FH},Y}(A):=
		\e^{\iu tH_{\mathrm{FH},Y}}A\e^{-\iu tH_{\mathrm{FH},Y}},
		\qquad
		\tau_t^{{\rm f},Y}(A):=\e^{\iu tH_{{\rm f},Y}}A\e^{-\iu tH_{{\rm f},Y}}.
	\end{equation}

	\begin{lemma}[Hubbard localization]\label{lem:Hubbard-localization}
		Under the finite-range condition and \eqref{eq:Qstar-Hubbard}, there are constants \(C,v>0\), depending only on \(D,R_*\), and \(Q_*\), but not on \(\Lambda\) or on the values of \(U_x\), such that for \(A\in\A_X\), \(t\geq0\), and \(R\geq1\),
		\begin{equation}\label{eq:Hubbard-localization}
			\norm{\tauFH_t(A)-\tau_t^{\mathrm{FH},X[R]}(A)}
			\leq CQ_*\norm A\,|X|(1+R)^{D-1}
			\e^{vt- R}.
		\end{equation}
		The analogous estimate holds for the free dynamics.
	\end{lemma}
	
	\begin{proof}
		The argument is essentially the same as in the proof of Step (I) for Theorem \ref{thm:transport-stability} and is thus kept very brief. We first establish the LRB as in the proof of Lemma \ref{lm:fermion-LR}:  As above, passing to the interaction picture removes the $U_x$-dependence in the on-site term $\sum_xU_xn_{x,\uparrow}n_{x,\downarrow}$, which, subsequently, allows to apply the standard fermionic LRB \cite[Theorem 3.1]{NachtergaeleSimsYoung2018Fermions}. Afterwards, we apply Duhamel to interpolate between the global and the local dynamics and apply the LRB.
	\end{proof}

	\smallskip\noindent\emph{Step (II): Local comparison.} For \(Z\subset\Lambda\), define
	\begin{equation}
		P_Z:=\prod_{x\in Z}(\one-n_{x,\downarrow}).
	\end{equation}
	Thus \(P_Z\) projects onto the sector with no spin-down particle in \(Z\).
	
	\begin{lemma}[Spin-down vacuum reduction]\label{lem:Hubbard-vacuum}
		Let \(X\subset Y\subset Z\subset\Lambda\), and let \(A\in\A_X\) satisfy \eqref{eq:down-preserving}. For every \(t\geq0\), \(P_Z\) commutes with \(\tau_t^{\mathrm{FH},Y}(A)\) and \(\tau_t^{{\rm f},Y}(A)\), and
		\begin{equation}\label{eq:projected-dynamics}
			P_Z\tau_t^{\mathrm{FH},Y}(A)=P_Z\tau_t^{{\rm f},Y}(A).
		\end{equation}
	\end{lemma}
	
	\begin{proof}
		The projection $P_Z$ reduces both restricted Hamiltonians. Since $Y\subset Z$,
		\begin{equation}
			P_Z(T_\downarrow)_Y=0,
			\qquad
			P_Z\sum_{x\in Y}U_xn_{x,\uparrow}n_{x,\downarrow}=0,
		\end{equation}
		and therefore
		\begin{equation}\label{eq:projected-H}
			P_ZH_{\mathrm{FH},Y}
			=
			P_ZH_{{\rm f},Y}
			=
			P_Z(T_\uparrow)_Y.
		\end{equation}
		
		Also, $A$ commutes with $P_Z$. Since $X\subset Z$, the projection factorizes as
		\begin{equation}
			P_Z=P_XP_{Z\setminus X}.
		\end{equation}
		The projection $P_X$ is the spectral projection of $N_{X,\downarrow}$ corresponding to the eigenvalue zero. Hence \eqref{eq:down-preserving} and functional calculus imply
		\begin{equation}
			[A,P_X]=0.
		\end{equation}
		On the other hand, $P_{Z\setminus X}$ is an even observable supported outside $X$, so locality gives
		\begin{equation}
			[A,P_{Z\setminus X}]=0.
		\end{equation}
		Consequently,
		\begin{align}
			[A,P_Z]
			&=[A,P_X]P_{Z\setminus X}
			+P_X[A,P_{Z\setminus X}]=0.
		\end{align}
		
		Together with
		\begin{equation}
			[P_Z,H_{\mathrm{FH},Y}]
			=
			[P_Z,H_{{\rm f},Y}]
			=
			0,
		\end{equation}
		this shows that $P_Z$ commutes with both restricted evolved observables. Functional calculus in the reducing subspace gives
		\begin{equation}\label{eq:projected-unitaries}
			P_Z\e^{\iu tH_{\mathrm{FH},Y}}
			=
			P_Z\e^{\iu tH_{{\rm f},Y}}
			=
			P_Z\e^{\iu t(T_\uparrow)_Y},
			\qquad t\geq0,
		\end{equation}
		which proves \eqref{eq:projected-dynamics}.
	\end{proof}

	Fix $R\geq1$. For each $j=1,\ldots,n$, define
	\begin{equation}
		Y_j:=X_j[R],
		\qquad
		Z:=\bigcup_{j=1}^nY_j.
	\end{equation}
	All products below are taken in the order assumed in \eqref{eq:Hubbard-main}. 
	
	We begin by replacing each full Fermi--Hubbard evolution with its restriction to $Y_j$. The telescoping identity reads
	\begin{align}
		&\prod_{j=1}^n\tauFH_{t_j}(A_j)
		-
		\prod_{j=1}^n\tau_{t_j}^{\mathrm{FH},Y_j}(A_j)
		\nonumber\\
		&\quad=
		\sum_{k=1}^n
		\left(
		\prod_{j<k}\tauFH_{t_j}(A_j)
		\right)
		\left(
		\tauFH_{t_k}(A_k)
		-
		\tau_{t_k}^{\mathrm{FH},Y_k}(A_k)
		\right)
		\left(
		\prod_{j>k}\tau_{t_j}^{\mathrm{FH},Y_j}(A_j)
		\right).
		\label{eq:Hubbard-telescope}
	\end{align}
	An empty product is understood to be the identity. Since the full and restricted evolutions are isometric, taking norms in \eqref{eq:Hubbard-telescope} gives
	\begin{align}
		&\norm{
			\prod_{j=1}^n\tauFH_{t_j}(A_j)
			-
			\prod_{j=1}^n\tau_{t_j}^{\mathrm{FH},Y_j}(A_j)
		}
		\nonumber\\
		&\quad\leq
		\sum_{k=1}^n
		\norm{
			\tauFH_{t_k}(A_k)
			-
			\tau_{t_k}^{\mathrm{FH},Y_k}(A_k)
		}
		\prod_{\substack{j=1\\j\neq k}}^n\norm{A_j}.
	\end{align}
	We may now apply Lemma~\ref{lem:Hubbard-localization} to every term in the sum. It follows that
	\begin{align}
		&\norm{
			\prod_{j=1}^n\tauFH_{t_j}(A_j)
			-
			\prod_{j=1}^n\tau_{t_j}^{\mathrm{FH},Y_j}(A_j)
		}
		\nonumber\\
		&\quad\leq
		CQ_*(1+R)^{D-1}\e^{-R}
		\left(
		\sum_{k=1}^n|X_k|\e^{vt_k}
		\right)
		\prod_{j=1}^n\norm{A_j}.
		\label{eq:Hubbard-product-localization}
	\end{align}
	
	The free dynamics satisfies the analogous localization estimate. Repeating the same telescoping argument gives
	\begin{align}
		&\norm{
			\prod_{j=1}^n\tauf_{t_j}(A_j)
			-
			\prod_{j=1}^n\tau_{t_j}^{{\rm f},Y_j}(A_j)
		}
		\nonumber\\
		&\quad\leq
		CQ_*(1+R)^{D-1}\e^{-R}
		\left(
		\sum_{k=1}^n|X_k|\e^{vt_k}
		\right)
		\prod_{j=1}^n\norm{A_j}.
		\label{eq:free-product-localization}
	\end{align}
	
	We next compare the two restricted evolutions. For every $j$, the inclusions
	\begin{equation}
		X_j\subset Y_j\subset Z
	\end{equation}
	allow us to apply Lemma~\ref{lem:Hubbard-vacuum}. Hence $P_Z$ commutes with both $\tau_{t_j}^{\mathrm{FH},Y_j}(A_j)$ and $\tau_{t_j}^{{\rm f},Y_j}(A_j)$, and we have
	\begin{equation}
		P_Z\tau_{t_j}^{\mathrm{FH},Y_j}(A_j)
		=
		P_Z\tau_{t_j}^{{\rm f},Y_j}(A_j).
	\end{equation}
	We can therefore move $P_Z$ through the ordered product and replace the factors one at a time. This yields
	\begin{equation}
		P_Z
		\prod_{j=1}^n\tau_{t_j}^{\mathrm{FH},Y_j}(A_j)
		=
		P_Z
		\prod_{j=1}^n\tau_{t_j}^{{\rm f},Y_j}(A_j).
		\label{eq:projected-products}
	\end{equation}

	Therefore,
	\begin{equation}
		P_Z\left(\prod_{j=1}^n\tau_{t_j}^{\mathrm{FH},Y_j}(A_j)-\prod_{j=1}^n\tau_{t_j}^{{\rm f},Y_j}(A_j)\right)=0=\left(\prod_{j=1}^n\tau_{t_j}^{\mathrm{FH},Y_j}(A_j)-\prod_{j=1}^n\tau_{t_j}^{{\rm f},Y_j}(A_j)\right)P_Z.
	\end{equation}
	And,
	\begin{equation}
		\begin{aligned}
			&\prod_{j=1}^n\tau_{t_j}^{\mathrm{FH},Y_j}(A_j)-\prod_{j=1}^n\tau_{t_j}^{{\rm f},Y_j}(A_j)\\
			&\quad=(\one-P_Z)\left(\prod_{j=1}^n\tau_{t_j}^{\mathrm{FH},Y_j}(A_j)-\prod_{j=1}^n\tau_{t_j}^{{\rm f},Y_j}(A_j)\right)(\one-P_Z).
		\end{aligned}
	\end{equation}
	
	By the cyclicity of the trace and the positivity of  $(\one-P_Z)\rho(\one-P_Z)$, 
	
	\begin{align}
		&\abs{\Tr\left(\rho \left(\prod_{j=1}^n\tau_{t_j}^{\mathrm{FH},Y_j}(A_j)-\prod_{j=1}^n\tau_{t_j}^{{\rm f},Y_j}(A_j)\right)\right)}\\
		&=\abs{\Tr\left(\rho(\one-P_Z) \left(\prod_{j=1}^n\tau_{t_j}^{\mathrm{FH},Y_j}(A_j)-\prod_{j=1}^n\tau_{t_j}^{{\rm f},Y_j}(A_j)\right)(\one-P_Z)\right)}\\
		&=\abs{\Tr\left((\one-P_Z)\rho(\one-P_Z) \left(\prod_{j=1}^n\tau_{t_j}^{\mathrm{FH},Y_j}(A_j)-\prod_{j=1}^n\tau_{t_j}^{{\rm f},Y_j}(A_j)\right)\right)}\\
		&\leq \norm{\prod_{j=1}^n\tau_{t_j}^{\mathrm{FH},Y_j}(A_j)-\prod_{j=1}^n\tau_{t_j}^{{\rm f},Y_j}(A_j)}\Tr ((\one-P_Z)\rho (\one-P_Z))\\
		&\leq 2\prod_{j=1}^n\norm{A_j}\Tr(\rho(\one-P_Z)).
	\end{align}
	
	Since $P_Z$ projects onto the sector with no spin-down particles in $Z$, we have $\one-P_Z\leq N_{Z,\downarrow}$. Moreover, $Z=\bigcup_{j=1}^nX_j[R]$. Assumption~\ref{ass:low-density} therefore implies
	\begin{align}
		\Tr\bigl(\rho(\one-P_Z)\bigr)
		&\leq
		\Tr\bigl(\rho N_{Z,\downarrow}\bigr)
		\leq
		\sum_{j=1}^n\sum_{x\in X_j}
		\Tr\bigl(\rho N_{\{x\}[R],\downarrow}\bigr)
		\nonumber\\
		&\leq
		\zeta(1+R)^D\sum_{j=1}^n|X_j|
		=
		\zeta S_X(1+R)^D.
		\label{eq:vacuum-weight}
	\end{align}
	Possible overlaps between the sets $\{x\}[R]$ only make the second estimate in the first line an overestimate. In conclusion, 
	\begin{equation}
		\abs{\Tr\left(\rho \left(\prod_{j=1}^n\tau_{t_j}^{\mathrm{FH},Y_j}(A_j)-\prod_{j=1}^n\tau_{t_j}^{{\rm f},Y_j}(A_j)\right)\right)}\leq 2\prod_{j=1}^n\norm{A_j}\zeta S_X(1+R)^D.
	\end{equation}

	We now combine the localization and no-spin-down comparisons. The triangle inequality yields
	\begin{align}
		&
		\abs{
			\Tr\left(
			\rho\prod_{j=1}^n\tauFH_{t_j}(A_j)
			\right)
			-
			\Tr\left(
			\rho\prod_{j=1}^n\tauf_{t_j}(A_j)
			\right)
		}
		\nonumber\\
		&\leq
		\abs{
			\Tr\left[
			\rho\left(
			\prod_{j=1}^n\tauFH_{t_j}(A_j)
			-
			\prod_{j=1}^n
			\tau_{t_j}^{\mathrm{FH},Y_j}(A_j)
			\right)
			\right]
		}
		\nonumber\\
		&\quad+
		\abs{
			\Tr\left[
			\rho \left(
			\prod_{j=1}^n
			\tau_{t_j}^{\mathrm{FH},Y_j}(A_j)
			-
			\prod_{j=1}^n
			\tau_{t_j}^{{\rm f},Y_j}(A_j)
			\right)
			\right]
		}
		\nonumber\\
		&\quad+
		\abs{
			\Tr\left[
			\rho\left(
			\prod_{j=1}^n
			\tau_{t_j}^{{\rm f},Y_j}(A_j)
			-
			\prod_{j=1}^n\tauf_{t_j}(A_j)
			\right)
			\right]
		}.
		\label{eq:Hubbard-three-errors}
	\end{align}
	The second term is bounded by estimate above. The first and last terms are controlled by \eqref{eq:Hubbard-product-localization} and \eqref{eq:free-product-localization}. After absorbing the resulting factor of two into $C$, we obtain \eqref{eq:Hubbard-main}.
	
	It remains to prove the polynomial-time conclusion. Fix the number and supports of the observables, assume that $0\leq t_j\leq T$ for every $j$, and choose
	\begin{equation}
		R=1+2vT.
	\end{equation}
	The localization term in \eqref{eq:Hubbard-main} then satisfies
	\begin{align}
		&Q_*(1+R)^{D-1}\e^{-R}
		\sum_{j=1}^n|X_j|\e^{vt_j}
		\nonumber\\
		&\quad\leq
		C(1+T)^{D-1}\e^{-1-2vT}\e^{vT}
		\nonumber\\
		&\quad\leq
		C(1+T)^{D-1}\e^{-vT}.
		\label{eq:Hubbard-window-LR}
	\end{align}
	Here the fixed model parameters, together with the number and supports of the observables, have been absorbed into $C$. The density term obeys
	\begin{align}
		\zeta S_X(1+R)^D
		&\leq
		C\zeta\,(1+T)^{D}.
		\label{eq:Hubbard-window-density}
	\end{align}
	Inserting \eqref{eq:Hubbard-window-LR} and \eqref{eq:Hubbard-window-density} into \eqref{eq:Hubbard-main} proves \eqref{eq:Hubbard-window}.
	
	Finally, let $T=\zeta^{-\alpha}$ with $0<\alpha<1/D$. The localization contribution satisfies
	\begin{equation}
		(1+\zeta^{-\alpha})^{D-1}
		\e^{-v\zeta^{-\alpha}}
		\longrightarrow0
		\qquad
		\text{as $\zeta\downarrow0$},
	\end{equation}
	because the exponential decay dominates every polynomial. At the same time,
	\begin{align}
		\zeta\,(1+\zeta^{-\alpha})^{D}
		&\leq
		C\zeta^{(1-\alpha D)}
		\longrightarrow0
		\qquad
		\text{as $\zeta\downarrow0$}.
	\end{align}
	The condition $\alpha<1/D$ is exactly what makes the exponent in the last expression positive. This completes the proof. \qed

	\subsection{Application to OTOCs} \label{subsec:OTOC}
	
	For $x,y\in\Lambda$ and $t\geq0$, define the interacting graded field OTOC of the spin-up creation/annihilation operators by
	\begin{equation}\label{eq:graded-OTOC}
		C_{xy,\mathrm{FH}}^{\Lambda}(t)
		:=
		\Tr\!\left(
		\rho\,
		\ac{a_{x,\uparrow}}{\tauFH_t(a_{y,\uparrow}^*)}^*
		\ac{a_{x,\uparrow}}{\tauFH_t(a_{y,\uparrow}^*)}
		\right).
	\end{equation}
	Its free counterpart is
	\begin{equation}
		C_{xy,{\rm f}}^\Lambda(t)
		:=
		\Tr\!\left(
		\rho\,
		\ac{a_{x,\uparrow}}{\tauf_t(a_{y,\uparrow}^*)}^*
		\ac{a_{x,\uparrow}}{\tauf_t(a_{y,\uparrow}^*)}
		\right).
	\end{equation}
	
	Let $h_{\uparrow,\Lambda}$ be the one-particle operator on $\ell^2(\Lambda)$ with matrix elements $J_{xy,\uparrow}$. Since $T_\uparrow$ is the second quantization of $h_{\uparrow,\Lambda}$, the free evolution of a creation operator is
	\begin{equation}
		\tauf_t(a_{y,\uparrow}^*)
		=
		\sum_{z\in\Lambda}
		\ip{\delta_z}{\e^{\iu t h_{\uparrow,\Lambda}}\delta_y}
		a_{z,\uparrow}^*.
	\end{equation}
	The CAR therefore give
	\begin{equation}
		\ac{a_{x,\uparrow}}{\tauf_t(a_{y,\uparrow}^*)}
		=
		\ip{\delta_x}{\e^{\iu t h_{\uparrow,\Lambda}}\delta_y}\one.
		\label{eq:free-graded-field}
	\end{equation}
	Thus the free OTOC is independent of $\rho$ and equals
	\begin{equation}\label{eq:free-graded-OTOC-finite}
		C_{xy,{\rm f}}^\Lambda(t)
		=
		\abs{
			\ip{\delta_x}
			{\e^{\iu t h_{\uparrow,\Lambda}}\delta_y}
		}^2.
	\end{equation}
	
	The following result, comparing the Fermi-Hubbard OTOC with its free counterpart, is a corollary to Theorem \ref{thm:Hubbard}. 
	
	\begin{corollary}[Graded field OTOC comparison]\label{cor:OTOC}
		Under the assumptions of Theorem~\ref{thm:Hubbard}, there are constants $C,v>0$, independent of $x,y,\Lambda,\rho$, and the values of $U_x$, such that, for every $t\geq0$ and $R\geq1$,
		\begin{equation}
			\abs{
				C_{xy,\mathrm{FH}}^{\Lambda}(t)
				-
				C_{xy,{\rm f}}^\Lambda(t)
			}
			\leq
			C\left(
			\e^{vt-R/2}
			+
			\zeta(1+R)^D
			\right).
			\label{eq:OTOC-bound}
		\end{equation}
	\end{corollary}
	
	\begin{proof}
		Expanding the positive operator in \eqref{eq:graded-OTOC} gives
		\begin{align}
			&
			\ac{a_{x,\uparrow}}{\tauFH_t(a_{y,\uparrow}^*)}^*
			\ac{a_{x,\uparrow}}{\tauFH_t(a_{y,\uparrow}^*)}
			\nonumber\\
			&=
			\tauFH_t(a_{y,\uparrow})
			a_{x,\uparrow}^*a_{x,\uparrow}
			\tauFH_t(a_{y,\uparrow}^*)
			\nonumber\\
			&\quad+
			\tauFH_t(a_{y,\uparrow})
			a_{x,\uparrow}^*
			\tauFH_t(a_{y,\uparrow}^*)
			a_{x,\uparrow}
			\nonumber\\
			&\quad+
			a_{x,\uparrow}^*
			\tauFH_t(a_{y,\uparrow})
			a_{x,\uparrow}
			\tauFH_t(a_{y,\uparrow}^*)
			\nonumber\\
			&\quad+
			a_{x,\uparrow}^*
			\tauFH_t(a_{y,\uparrow})
			\tauFH_t(a_{y,\uparrow}^*)
			a_{x,\uparrow}.
			\label{eq:graded-OTOC-expansion}
		\end{align}
		
		The local fields appearing in the expansion preserve the spin-down vacuum sector. More explicitly,
		\begin{align}
			[a_{x,\uparrow},N_{\{x\},\downarrow}]
			&=
			[a_{x,\uparrow}^*,N_{\{x\},\downarrow}]
			=0,
			\nonumber\\
			[a_{y,\uparrow},N_{\{y\},\downarrow}]
			&=
			[a_{y,\uparrow}^*,N_{\{y\},\downarrow}]
			=0.
		\end{align}
		Theorem~\ref{thm:Hubbard} can therefore be applied separately to the expectation of each term in \eqref{eq:graded-OTOC-expansion}. In every case there are four single-site observables, two evaluated at time $0$ and two at time $t$. Hence $S_X=4$, and all operator norms are at most one.
		
		The localization part of \eqref{eq:Hubbard-main} is bounded by
		\begin{equation}
			CQ_*(1+R)^{D-1}\e^{-R}\bigl(2+2\e^{vt}\bigr)
			\leq
			C\e^{vt-R/2},
		\end{equation}
		where we used $(1+R)^{D-1}\e^{-R}\leq C\e^{-R/2}$ for $R\geq1$.
		The density contribution is bounded by $C\zeta(1+R)^D$. Summing the four comparisons and absorbing their number into $C$ proves \eqref{eq:OTOC-bound}.
	\end{proof}
	
	\subsection{Exact graded OTOC for the full-space lattice Laplacian}
	\label{sec:free-laplacian-OTOC}
	
	We now specialize Corollary~\ref{cor:OTOC} to nearest-neighbor hopping on the cubic lattice, i.e.~set $h_{\uparrow, \Lambda}$ to be the discrete Laplacian $\Delta_\Lambda$. However, although the finite-volume free OTOC is already a one-particle quantity, its explicit evaluation is inconvenient.
	We therefore compare it with the corresponding full-space propagator, which can be computed by Fourier transformation. This is Step~(III) of the comparison argument.
	
	On $\ell^2(\mathbb Z^D)$, let
	\begin{equation}\label{eq:free-graph-Laplacian}
		-\Delta_{\mathbb Z^D}f(x)
		=
		2Df(x)
		-
		\sum_{\substack{y\in\mathbb Z^D\\y\sim x}}f(y).
	\end{equation}
	We write $\Delta_\Lambda$ for the finite-volume restriction of $\Delta_{\mathbb Z^D}$, retaining the constant diagonal term $-2D$.
	For $t\geq0$, define
	\begin{equation}\label{eq:free-OTOC-kernel}
		C_{xy,{\rm f}}^\infty(t)
		:=
		\abs{
			\ip{\delta_x}
			{\e^{\iu t\Delta_{\mathbb Z^D}}\delta_y}
		}^2.
	\end{equation}

	\begin{lemma}[Finite- to full-space free comparison]
		\label{lem:free-finite-full}
		There are constants $C,v_0>0$, depending only on $D$, such that, whenever $x,y\in\Lambda$, $t\geq0$, and $B_b(y)\subset\Lambda$,
		\begin{equation}
			\abs{
				C_{xy,{\rm f}}^\Lambda(t)
				-
				C_{xy,{\rm f}}^\infty(t)
			}
			\leq
			C\e^{v_0t-b/2}.
			\label{eq:free-finite-full}
		\end{equation}
	\end{lemma}
	
	\begin{proof}
		Apply the one-particle boundary estimate \eqref{eq:onebody-boundary} to $\Delta_\Lambda$ and $\Delta_{\mathbb Z^D}$. After extending the finite-volume vector by zero, we obtain
		\begin{equation}\label{eq:free-kernel-finite-full}
			\abs{
				\ip{\delta_x}{\e^{\iu t\Delta_\Lambda}\delta_y}
				-
				\ip{\delta_x}
				{\e^{\iu t\Delta_{\mathbb Z^D}}\delta_y}
			}
			\leq
			C(1+b)^{D-1}\e^{v_0t-b}.
		\end{equation}
		The polynomial prefactor can be absorbed into half of the exponential:
		\begin{equation}
			(1+b)^{D-1}\e^{-b}
			\leq
			C\e^{-b/2},
			\qquad b\geq1.
		\end{equation}
		Finally, both propagator matrix elements have modulus at most one.
		Using
		\begin{equation}
			\bigl||z|^2-|w|^2\bigr|
			\leq
			2|z-w|,
			\qquad |z|,|w|\leq1,
		\end{equation}
		this proves \eqref{eq:free-finite-full}.
	\end{proof}
	In the following lemma, we evaluate the full-space free graded OTOC. 
	\begin{lemma}[Exact full-space free graded OTOC]
		\label{lem:free-OTOC-exact}
		For \(x,y\in\mathbb Z^D\), set
		\begin{equation}
			r_\mu:=|x_\mu-y_\mu|,
			\qquad
			\ell:=\sum_{\mu=1}^D r_\mu=|x-y|_1.
		\end{equation}
		Then, for every \(t\geq0\),
		\begin{equation}\label{eq:free-OTOC-Bessel}
			C_{xy,{\rm f}}^\infty(t)
			=\prod_{\mu=1}^D J_{r_\mu}(2t)^2,
		\end{equation}
		where $J_m$ denotes the Bessel function of the first kind of order $m$.
		For fixed \(x,y\), its small-time expansion is
		\begin{equation}\label{eq:free-OTOC-small-time}
			C_{xy,{\rm f}}^\infty(t)=
			\frac{t^{2\ell}}{\prod_{\mu=1}^D(r_\mu!)^2}
			\left[
			1-2t^2\sum_{\mu=1}^D\frac{1}{r_\mu+1}+O(t^4)
			\right].
		\end{equation}
		Equivalently, if
		\begin{equation}
			N_{\mathrm{sp}}(x,y):=\frac{\ell!}{r_1!\cdots r_D!}
		\end{equation}
		is the number of shortest lattice paths from \(y\) to \(x\), then
		\begin{equation}\label{eq:free-OTOC-paths}
			C_{xy,{\rm f}}^\infty(t)
			=\left(\frac{N_{\mathrm{sp}}(x,y)}{\ell!}\right)^2t^{2\ell}
			+O(t^{2\ell+2}).
		\end{equation}
	\end{lemma}
	
	\begin{proof} For every integer $m\geq0$, we use the integral representation
		\begin{equation}\label{eq:Bessel-integral}
			\iu^mJ_m(2t)
			=
			\frac{1}{2\pi}
			\int_{-\pi}^{\pi}
			\e^{\iu m\theta}\e^{2\iu t\cos\theta}\dd\theta.
		\end{equation}
		Fourier transformation gives
		\begin{align}
			\ip{\delta_x}
			{\e^{\iu t\Delta_{\mathbb Z^D}}\delta_y}
			&=
			\frac{1}{(2\pi)^D}
			\int_{[-\pi,\pi]^D}
			\e^{\iu k\cdot(x-y)}
			\e^{2\iu t\sum_{\mu=1}^D(\cos k_\mu-1)}
			\dd k
			\nonumber\\
			&=
			\e^{-2D\iu t}
			\prod_{\mu=1}^D
			\left[
			\frac{1}{2\pi}
			\int_{-\pi}^{\pi}
			\e^{\iu k_\mu(x_\mu-y_\mu)}
			\e^{2\iu t\cos k_\mu}
			\dd k_\mu
			\right]
			\nonumber\\
			&=
			\e^{-2D\iu t}\iu^\ell
			\prod_{\mu=1}^D J_{r_\mu}(2t).
		\end{align}
		In the last step we used \eqref{eq:Bessel-integral} together with $J_{-m}(z)=(-1)^mJ_m(z)$. Taking the squared modulus removes the phase and proves \eqref{eq:free-OTOC-Bessel}.
		
		The usual power series expansion for the Bessel function gives
		\begin{align}
			J_m(2t)
			=
			\sum_{q=0}^\infty
			\frac{(-1)^q t^{m+2q}}{q!(m+q)!}
			=
			\frac{t^m}{m!}
			\left(
			1-\frac{t^2}{m+1}+O(t^4)
			\right).
		\end{align}
		Multiplying these expansions over the $D$ coordinates proves \eqref{eq:free-OTOC-small-time}. Finally,
		\begin{equation}
			\frac{\ell!}{r_1!\cdots r_D!}
		\end{equation}
		is the multinomial coefficient counting the shortest nearest-neighbor paths from $y$ to $x$. Substitution gives \eqref{eq:free-OTOC-paths}.
	\end{proof}

	\subsubsection{Approximation by the full-space hopping OTOC} Combining Corollary~\ref{cor:OTOC} with Lemma~\ref{lem:free-finite-full} and using the triangle inequality gives
	\begin{align}
		&
		\abs{
			C_{xy,\mathrm{FH}}^\Lambda(t)
			-
			C_{xy,{\rm f}}^\infty(t)
		}
		\nonumber\\
		&\quad\leq
		C\left(
		\zeta(1+R)^D
		+
		\e^{vt-R/2}
		+
		\e^{v_0t-b/2}
		\right).
		\label{eq:interacting-Bessel-comparison}
	\end{align}
	The three terms have different origins. The parameter $b$ measures the distance available between the initial site $y$ and the boundary of $\Lambda$. Thus the finite-volume error is small whenever $b$ is large compared with $2v_0t$. Similarly, taking $R$ larger than $2vt$ makes the Hubbard localization error small. The remaining contribution, $\zeta(1+R)^D$, bounds the expected number of spin-down particles inside each of the balls of radius $R$ centered at $x$ and $y$.
	
	\subsubsection{The knight beats the rook}
	We now compare sites at the same graph distance. For $D\geq2$, \eqref{eq:free-OTOC-paths} gives
	\begin{equation}
		C_{(2,0,\ldots),0,{\rm f}}^\infty(t)
		=
		\frac{t^4}{4}+O(t^6),
		\qquad
		C_{(1,1,0,\ldots),0,{\rm f}}^\infty(t)
		=
		t^4+O(t^6).
	\end{equation}
	Both sites are at graph distance two from the origin. The diagonal site, however, can be reached by two shortest paths, whereas the axial site can be reached by only one. Their amplitudes add coherently, so the leading OTOC is larger by a factor of four.
	
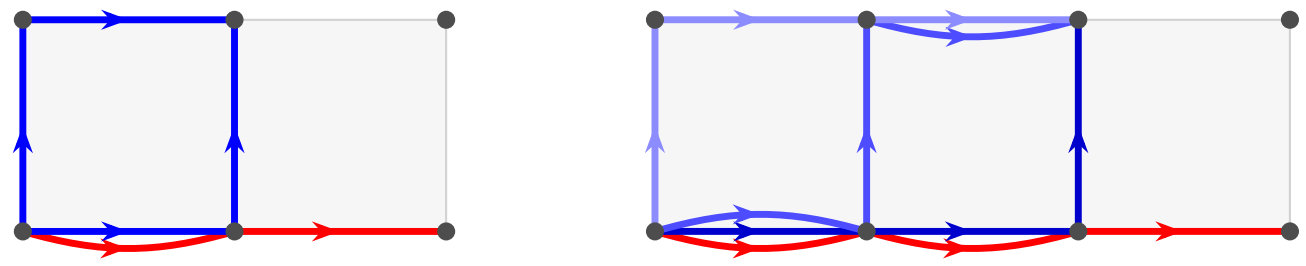
\begin{figure}
		\centering
		\begin{tikzpicture}[
			lattice picture,
			shortest path/.append style={
				path arrows={0.25}{0.5}
			}
			]
			\latticebackground{2}
			
			\draw[red path]
			(0,0) to[bend right=16] (1,0)
			-- (2,0);
			
			\draw[blue path]
			(0,0) -- (1,0) -- (1,1);
			
			\draw[blue path]
			(0,0) -- (0,1) -- (1,1);
			
			\latticevertices{2}
		\end{tikzpicture}
		\hspace{1cm}
		\begin{tikzpicture}[
			lattice picture,
			shortest path/.append style={
				path arrows={0.166667}{0.333333}
			}
			]
			\latticebackground{3}
			
			\draw[red path]
			(0,0) to[bend right=16] (1,0)
			to[bend right=16] (2,0)
			-- (3,0);
			
			\draw[blue dark]
			(0,0) -- (1,0) -- (2,0) -- (2,1);
			
			\draw[blue medium]
			(0,0) to[bend left=16] (1,0)
			-- (1,1)
			to[bend right=16] (2,1);
			
			\draw[blue light]
			(0,0) -- (0,1) -- (1,1) -- (2,1);
			
			\latticevertices{3}
		\end{tikzpicture}  
		\caption{Shortest paths of lengths two and three.}
		\label{fig:OTOC}
	\end{figure}

	The same combinatorial effect becomes more pronounced at graph distance three:
	\begin{equation}
		C_{(2,1,0,\ldots),0,{\rm f}}^\infty(t)
		=
		\frac{t^6}{4}+O(t^8),
		\qquad
		C_{(3,0,\ldots),0,{\rm f}}^\infty(t)
		=
		\frac{t^6}{36}+O(t^8).
	\end{equation}
	There are three shortest paths to $(2,1,0,\ldots)$ and only one to $(3,0,\ldots)$; cf.~Figure \ref{fig:OTOC}. Squaring the corresponding ratio of amplitudes gives the factor nine. This is the path-counting mechanism sometimes described as ``the knight beats the rook'', because (in chess) the knight can arrive at its destination by more than one path while the rook only has one path available \cite[Figure 5]{bippus2026two}.
	
	The comparison estimate \eqref{eq:interacting-Bessel-comparison} transfers this single-particle pattern to the interacting graded field OTOC whenever its right-hand side is $o(t^{2\ell})$. Under that condition, the interacting OTOC inherits both the graph-distance exponent and the squared shortest-path multiplicity.
	
	Our rigorous statements concern operator spreading, as measured by the graded field OTOC, rather than two-site entanglement itself. Nevertheless, they support the following heuristic picture: at short times, two-site entanglement spreading in Fermi-Hubbard models is governed primarily by the non-interacting hopping dynamics, even when the interaction is not small. This interpretation is consistent with the numerical results in \cite[Figure~4]{bippus2026two}, obtained at interaction strength $U_x=2$. In this sense, our result provides a rigorous OTOC counterpart to the  entanglement distribution observed  in those simulations.

	\appendix
	
	\section{Lattice estimates}\label{sec:tails}
	
	We record the elementary bounds used above.  Their constants are deliberately not optimized.
	
	\begin{lemma}[Exponential lattice tails]\label{lem:lattice-tails}
		For \(D\in\mathbb N\), \(\alpha>0\), and \(m\geq0\), there are \(C=C(D,m,\alpha)\) and \(R_0=R_0(D,m,\alpha)\) such that
		\begin{align}
			\sum_{\substack{x\in\mathbb Z^D\\|x|\leq R}}|x|^m
			&\leq C(1+R)^{D+m},\label{eq:inner-lattice}\\
			\sum_{\substack{x\in\mathbb Z^D\\|x|>R}}|x|^m\e^{-\alpha|x|}
			&\leq C(1+R)^{D+m-1}\e^{-\alpha R},
			\qquad R\geq R_0.\label{eq:outer-lattice}
		\end{align}
	\end{lemma}
	
	\begin{proof}
		The number of lattice points in the shell \(S_k:=\{x:k\leq|x|<k+1\}\) is bounded by
		\begin{equation}\label{eq:shell-count}
			|S_k|\leq C_D(1+k)^{D-1}.
		\end{equation}
		Therefore
		\begin{align}
			\sum_{|x|\leq R}|x|^m
			&\leq C_D\sum_{0\leq k\leq\lceil R\rceil}
			(1+k)^{D+m-1}\nonumber\\
			&\leq C_{D,m}(1+R)^{D+m},
		\end{align}
		which proves \eqref{eq:inner-lattice}.  Similarly,
		\begin{align}
			\sum_{|x|>R}|x|^m\e^{-\alpha|x|}
			&\leq C_{D,m,\alpha}
			\sum_{k\geq\lfloor R\rfloor}(1+k)^{D+m-1}\e^{-\alpha k}\nonumber\\
			&\leq C_{D,m,\alpha}
			\int_R^\infty(1+s)^{D+m-1}\e^{-\alpha s}\dd s. \label{eq:tail-integral}
		\end{align}
		Increasing the constant handles the bounded shift between the series and the integral.  Put \(p=D+m-1\).  For \(R\geq R_0:=2p/\alpha\), the function \((1+s)^p\e^{-\alpha s/2}\) is decreasing on \([R,\infty)\), and hence
		\begin{align}
			\int_R^\infty(1+s)^p\e^{-\alpha s}\dd s
			&=\int_R^\infty
			\left((1+s)^p\e^{-\alpha s/2}\right)
			\e^{-\alpha s/2}\dd s\nonumber\\
			&\leq(1+R)^p\e^{-\alpha R/2}
			\int_R^\infty\e^{-\alpha s/2}\dd s\nonumber\\
			&=\frac2\alpha(1+R)^p\e^{-\alpha R}.
		\end{align}
		Together with \eqref{eq:tail-integral}, this proves \eqref{eq:outer-lattice} for every real \(m\geq0\).
	\end{proof}
	
	\section*{Acknowledgment}	
	The research of M.L.\ is supported by the DFG through the grants TRR 352 – Project-ID 470903074 and FOR 5413 - Project-ID 465199066. The research of M.L., J.A.F., and D.T.\ is supported by the European Union through ERC Starting Grant MathQuantProp, Grant Agreement 101163620.\footnote{Views and opinions expressed are however those of the authors only and do not necessarily reflect those of the European Union or the European Research Council Executive Agency. Neither the European Union nor the granting authority can be held responsible for them.}
	J.H. acknowledges funding from the European Research Council (ERC) and the Swiss State Secretariat for Education, Research and Innovation (SERI) through the consolidator grant ProbQuant.
	\section*{Declarations}
	
	\paragraph{Data availability.}
	No datasets were generated or analyzed in this work. The authors acknowledge the use of AI tools for writing and streamlining the presentation. The proofs were discovered by the authors without AI.

	\bibliographystyle{plainurl} 
	\bibliography{biblio-updated}
\end{document}